\documentclass[11pt,a4paper,english,twoside]{article}

  \usepackage{a4wide}
  \usepackage{latexsym}
  \usepackage{epsf}
  \usepackage{amssymb}
  \usepackage{graphicx}
  \usepackage{cite}
  \usepackage{amsmath,amssymb,amsthm}
  \usepackage{verbatim}
  \usepackage{hyperref}
  \hypersetup{linktocpage}
  \usepackage[english]{babel}
  \usepackage{enumerate}
  \usepackage{dsfont}
  \usepackage{empheq}

\newcommand{\beq}{\begin{equation}}
\newcommand{\eeq}{\end{equation}}
\def\bea#1\eea{\begin{align}#1\end{align}}

\newcommand{\nn}{\nonumber}

\renewcommand{\d}{\textrm{d}}

\newcommand{\w}{\wedge}

\def\del {\partial}
\def\mmm {\mathcal{M}}
\def\NS {N\!S}
\def\KK {K\!K}

\begin{document}

\begin{titlepage}
  \begin{flushright}
  \small
  IPhT-t16/080
  \normalsize
  \end{flushright}

\begin{center}

\phantom{DRAFT}

\vspace{1.5cm}

{\LARGE \bf{Refining the boundaries \vspace{0.3cm}\\ of the classical de Sitter landscape}}\\

\vspace{2 cm} {\Large David Andriot$^{1,2}$ and Johan Bl{\aa}b{\"a}ck$^3$}\\
 \vspace{0.9 cm} {\small\slshape $^1$ Max-Planck-Institut f\"ur Gravitationsphysik, Albert-Einstein-Institut,\\Am M\"uhlenberg 1, 14467 Potsdam-Golm, Germany}\\
 \vspace{0.2 cm} {\small\slshape $^2$ Institut f\"ur Mathematik, Humboldt-Universit\"at zu Berlin, IRIS-Adlershof,\\Zum Gro\ss en Windkanal 6, 12489 Berlin, Germany}\\
 \vspace{0.2 cm} {\small\slshape $^3$ Institut de Physique Th\'eorique, Universit\'e Paris Saclay, CEA, CNRS,\\F-91191 Gif sur Yvette, France}\\
\vspace{0.5cm} {\upshape\ttfamily david.andriot@aei.mpg.de, johan.blaback@cea.fr}\\

\vspace{3cm}

{\bf Abstract}
\end{center}

\begin{quotation}
We derive highly constraining no-go theorems for classical de Sitter backgrounds of string theory, with parallel sources; this should impact the embedding of cosmological models. We study ten-dimensional vacua of type II supergravities with parallel and backreacted orientifold $O_p$-planes and $D_p$-branes, on four-dimensional de Sitter space-time times a compact manifold. Vacua for $p=3$, $7$ or $8$ are completely excluded, and we obtain tight constraints for $p=4$, $5$, $6$. This is achieved through the derivation of an enlightening expression for the four-dimensional Ricci scalar. Further interesting expressions and no-go theorems are obtained. The paper is self-contained so technical aspects, including conventions, might be of more general interest.
\end{quotation}

\end{titlepage}

\newpage

\pagenumbering{gobble}

\begin{center}
{\LARGE \bf{Erratum}}\\
\end{center}

\begin{itemize}
  \item \bf{Summary and consequences}
\end{itemize}

\noindent In the trace of the Einstein equation along internal parallel flat directions, namely equations \eqref{traceparIIA} and \eqref{traceparIIB}, a few terms have been missed. As a consequence the corrected equations will have additional terms which depend on specific components of fluxes, such as $|H^{(2)}|^2 + 2 |H^{(3)}|^2$ which are the squares of the components $H_{a_{||}b_{||}c_{\bot}}$ and $H_{a_{||}b_{||}c_{||}}$. These terms are then absent in the final ${\tilde{\cal R}}_4$ expression \eqref{FINAL} and \eqref{FINALint}. The only change that impacts our conclusion, \eqref{constraintcurv}, is that the curvature terms $2 {\cal R}_{||} + 2 {\cal R}_{||}^{\bot}$ should be replaced by
\begin{equation}
  2 {\cal R}_{||} + 2 {\cal R}_{||}^{\bot} - |H^{(2)}|^2 - 2|H^{(3)}|^2 \ . \label{combifields}
\end{equation}
As in \eqref{constraintcurv}, this combination gets bounded by two inequalities, in order to get classical de Sitter solutions for parallel $p=4,5,6$ sources. While this change modifies the final expression, it has little impact on the physics result: we obtain tight constraints on a combination of fields for de Sitter solutions to exist with parallel $p=4,5,6$ sources. The no-go theorems for parallel $p=3,7,8$ sources are not affected at all.

The combination \eqref{combifields} is better motivated than the curvature terms alone, as it now appears to be T-duality invariant, on geometric backgrounds. This statement can be made more precise by considering group manifolds, where the $f^a{}_{bc}$, building the curvature terms, are constant, and some are set to zero by the orientifold projection. In addition, the $H$-flux is odd under an orientifold involution, imposing $H^{(3)}=0$ for a constant flux; avoiding the Freed-Witten anomaly also sets $H^{(3)}$ to zero. The (opposite sign of the) combination \eqref{combifields} then reduces to
\begin{equation}
  \begin{split}\raisetag{21pt}
    & \delta^{ab} f^{d_{||}}{}_{c_{||}a_{||}} f^{c_{||}}{}_{d_{||}b_{||}} + \frac{1}{2} \delta^{ch}\delta^{dj}\delta_{ab} f^{a_{||}}{}_{c_{||}j_{||}} f^{b_{||}}{}_{h_{||}d_{||}} + \delta^{ab} f^{d_{\bot}}{}_{c_{\bot}a_{||}} f^{c_{\bot}}{}_{d_{\bot}b_{||}} + \delta^{ab} \delta^{dg} \delta_{ch} f^{h_{\bot}}{}_{g_{\bot}a_{||}} f^{c_{\bot}}{}_{d_{\bot}b_{||}}\\
    & + \frac{1}{2} \delta^{ad}\delta^{be}\delta^{cf} H_{a_{||}b_{||}c_{\bot}} H_{d_{||}e_{||}f_{\bot}}\ ,\label{newqtty2}
  \end{split}
\end{equation}
and the first and third terms vanish on nilmanifolds. The $H$-flux component schematically transforms under T-duality into one or the other structure constant, depending on the T-duality direction
\begin{equation}
  H_{a_{||}b_{||}c_{\bot}} \rightarrow f^{c_{||}}{}_{a_{||}b_{||}} \ {\rm or}\ -f^{a_{\bot}}{}_{c_{\bot}b_{||}} \ ,
\end{equation}
showing the T-duality invariance of the combination \eqref{combifields} in that setting.

A practical consequence for the paper is that several occurrences of ``curvature terms'' should be replaced by the above ``field combination'': it is the case for equations \eqref{nongo456intro}, \eqref{FINALintro}, \eqref{nongo456}, and the text of the Outlook. The discussed consequences of the results are unchanged: to start with, the remark on the solutions T-dual to one with an $O_3$, at the end of Section \ref{sec:nogo3456}, remains valid. The requirement of having $f^{a_{||}}{}_{b_{\bot}c_{\bot}} \neq 0$ for a de Sitter solution still holds, from the constraints on the new combination, implying the no-go theorem for $p=8$ (Footnote \ref{foot:p=8}) and the impossibility to embed a specific monodromy inflation mechanism, as mentioned in the Outlook.

\begin{itemize}
  \item \bf{Corrected equations}
\end{itemize}

\noindent For a $p$-dimensional source, any internal flux $F_q$ was decomposed in \eqref{decompoflux} as $F_q=\sum_{n=0}^{p-3} F_q^{(n)}$, where the components of $F_q^{(n)}$ have $n$ internal parallel flat indices, and $F_q^{(0)}=F_q|_{\bot}$. As a consequence, one has
\begin{equation}
  \begin{split}
    & |F_q|^2= \sum_{n=0}^{p-3} |F_q^{(n)}|^2 \ ,\ {\rm where}\ |F_q|^2= \frac{1}{q!} F_{q\ a_1 \dots a_q} F_q^{a_1 \dots a_q} \ , \label{sumsquareflux}\\
    & |F_q^{(n)}|^2= \frac{1}{n!(q-n)!} F_{q\ a_{1||} \dots a_{n||}a_{n+1\bot} \dots a_{q\bot}} F_q^{a_{1||} \dots a_{n||}a_{n+1\bot} \dots a_{q\bot}} \ ,
  \end{split}
\end{equation}
the indices being lifted by the flat internal metric. We now consider the trace of the Einstein equation along the internal parallel directions. An internal flux $F_q$ appears in it as follows
\begin{equation}
  \begin{split}\raisetag{26pt}
    & \delta^{ab} \frac{1}{(q-1)!} F_{q\ a_{1||} a_2 \dots a_q} F_{q\ b_{1||}}^{\ \ \ \ a_2 \dots a_q} = \sum_{n\geq 1}^{p-3} \delta^{ab} \frac{1}{(n-1)!(q-n)!} F^{(n)}_{q\ a_{1||} a_{2||} \dots a_{n||}a_{n+1\bot} \dots a_{q\bot}} F_{q\ \ b_{1||}}^{(n) \ \ a_{2||} \dots a_{n||}a_{n+1\bot} \dots a_{q\bot}}\\
    & = \sum_{n\geq 0}^{p-3} n |F_q^{(n)}|^2 = |F_q|^2 - |F_q|_{\bot}|^2 + \sum_{n\geq 2}^{p-3} (n-1) |F_q^{(n)}|^2 \ .
  \end{split}
\end{equation}
The last sum is absent of \eqref{traceparIIA} and \eqref{traceparIIB}. These two equations are corrected towards
\begin{equation}
  \begin{split}
    {\cal R}_{6||} + 2 (\nabla\del \phi)_{6||} &= \frac{p-3}{4} \left({\cal R}_4 +  2 (\nabla\del \phi)_4 + 2e^{2\phi} |F_{6}|^2  \right) \\
    & + \frac{1}{2} \left(|H|^2 - |H|_{\bot}|^2 + e^{2\phi} (  |F_{2}|^2 - |F_{2}|_{\bot}|^2 + |F_{4}|^2 - |F_{4}|_{\bot}|^2 \right) \\
    & + \frac{1}{2} \sum_{n\geq 2}^{p-3} (n-1) \left(|H^{(n)}|^2 + e^{2\phi} (  |F_2^{(n)}|^2 + |F_4^{(n)}|^2 ) \right) \\
    {\cal R}_{6||} + 2 (\nabla\del \phi)_{6||} &= \frac{p-3}{4} \left({\cal R}_4 +  2 (\nabla\del \phi)_4 + e^{2\phi} |F_{5}|^2  \right) \\
    & + \frac{1}{2} \left(|H|^2 - |H|_{\bot}|^2 + e^{2\phi} (  |F_{1}|^2 - |F_{1}|_{\bot}|^2 + |F_{3}|^2 - |F_{3}|_{\bot}|^2 \right) \\
    & + \frac{1}{4} e^{2\phi} \left(  |F_{5}|^2 - |F_{5}|_{\bot}|^2 - |*_6 F_{5}|^2 + |(*_6 F_{5})|_{\bot}|^2 \right)  \\
    & + \frac{1}{2} \sum_{n\geq 2}^{p-3} (n-1) \left(|H^{(n)}|^2 + e^{2\phi} (  |F_3^{(n)}|^2 + \frac{1}{2} |F_5^{(n)}|^2 ) \right) \ ,
  \end{split}
\end{equation}
where in IIB, the one-form fluxes, $F_1$ and $*_6 F_{5}$, do not contribute to the new terms because the sum starts with $n\geq 2$. For the same reason, these new terms only contribute for $p\geq 5$. A general rewriting of these two equations, correcting equation \eqref{tracepargen}, is then given by
\begin{equation}
  \begin{split}\raisetag{26pt}
    \hspace{-0.2in} 2 {\cal R}_{6||} + & 4 (\nabla\del \phi)_{6||} -  \frac{p-3}{2} \left({\cal R}_4  + 2 (\nabla\del \phi)_4 \right) = |H|^2 - |H|_{\bot}|^2 + e^{2\phi} \left(  |F_{k-2}|^2 - |F_{k-2}|_{\bot}|^2 \right) \label{tracepargenEr}\\
    & + e^{2\phi} \bigg(  |F_{k}|^2 - |F_{k}|_{\bot}|^2 + |F_{k+2}|^2 + (9-p) |F_{k+4}|^2 + 5 |F_{k+6}|^2 + \frac{1}{2} (|(*_6 F_{5})|_{\bot}|^2 - |F_{5}|_{\bot}|^2) \bigg) \\
    & + \sum_{n\geq 2}^{p-3} (n-1) \left(|H^{(n)}|^2 + e^{2\phi} (  |F_k^{(n)}|^2 + |F_{k+2}^{(n)}|^2 + \frac{p-6}{2} |F_{k+4}^{(n)}|^2 + \frac{p-7}{4} |F_5^{(n)}|^2 ) \right) \ ,
  \end{split}
\end{equation}
where the $F_{5}$ terms should only be considered in IIB. Equation \eqref{combine1} gets corrected by adding the same new line, while the final formula \eqref{FINAL} becomes
\begin{equation}
  \begin{split}
    & 2 e^{-2A} \tilde{{\cal R}}_4  = - \left|*_{\bot}H|_{\bot} + \varepsilon_p e^{\phi} F_{k-2}|_{\bot} \right|^2  - 2 e^{2\phi}\left| g_s^{-1} \tilde{*}_{\bot} \d e^{-4A} - \varepsilon_p  F_{k}^{(0)} \right|^2 \label{FINALEr} \\
    & \phantom{2 e^{-2A} \tilde{{\cal R}}_4  =} - \sum_{a_{||}} \left| *_{\bot}( \d e^{a_{||}})|_{\bot} - \varepsilon_p e^{\phi} (\iota_{\del_{a_{||}}} F_k^{(1)} ) \right|^2\ - 2 {\cal R}_{||} - 2 {\cal R}_{||}^{\bot}  \\
    & \phantom{2 e^{-2A} \tilde{{\cal R}}_4  =} - 2 e^{-2A} \left(\d \left(e^{8A} \tilde{*}_{\bot} \d e^{-4A} - e^{8A} \varepsilon_p g_s F_{k}^{(0)} \right)\right)_{\widetilde{\bot}} \\
    & \ - e^{2\phi} \Big( |F_{k}|^2 - |F_{k}^{(0)}|^2 - |F_{k}^{(1)}|^2 + 2 |F_{k+2}|^2 + (p-5) |F_{k+4}|^2  + \frac{1}{2} (|F_{5}|_{\bot}|^2 - |(*_6 F_{5})|_{\bot}|^2) \Big) \\
    & + \sum_{n\geq 2}^{p-3} (n-1) \left(|H^{(n)}|^2 + e^{2\phi} (  |F_k^{(n)}|^2 + |F_{k+2}^{(n)}|^2 + \frac{p-6}{2} |F_{k+4}^{(n)}|^2 + \frac{p-7}{4} |F_5^{(n)}|^2 ) \right)  \ .
  \end{split}
\end{equation}
We now detail the last two lines of \eqref{FINALEr}: they are equal to
\begin{equation}
  \begin{split}
    & p=3:\ 0 \\
    & p=4:\ - 2 e^{2\phi} |F_{6}|^2 \\
    & p=5:\ |H^{(2)}|^2 - e^{2\phi} \Big( 2 |F_{5}|^2 - \frac{1}{2} |(*_6 F_{5})|_{\bot}|^2 - \frac{1}{2} |F_{5}^{(2)}|^2 \Big) \\
    & p=6:\ |H^{(2)}|^2 + 2 |H^{(3)}|^2 - e^{2\phi} \Big( 2 |F_{4}|^2 - |F_{4}^{(2)}|^2 - 2 |F_{4}^{(3)}|^2 + |F_{6}|^2 \Big) \\
    & p=7:\ |H^{(2)}|^2 + 2 |H^{(3)}|^2 - e^{2\phi} \Big( 2 |F_{3}|^2 - |F_{3}^{(2)}|^2 - 2 |F_{3}^{(3)}|^2 + 2|F_{5}|^2 - \frac{1}{2} |(*_6 F_{5})|_{\bot}|^2\\
    & \phantom{p=7:\ |H^{(2)}|^2 + 2 |H^{(3)}|^2 - e^{2\phi} \Big( 2 |F_{3}|^2 - |F_{3}^{(2)}|^2 - 2 |F_{3}^{(3)}|^2}\ - \frac{1}{2} \sum_{n\geq 2}^{4} (n-1) |F_5^{(n)}|^2 \Big)\\
    & p=8:\ |H^{(2)}|^2 + 2 |H^{(3)}|^2 - e^{2\phi} \Big( 2 |F_{2}|^2 - |F_{2}^{(2)}|^2 + 3 |F_{4}|^2 - \sum_{n\geq 2}^{4} (n-1) |F_4^{(n)}|^2  \Big) \ .
  \end{split}
\end{equation}
We used \eqref{sumsquareflux}, that leads to the cancelation of all $F_k$ terms. That equation, together with $|F_{5}|^2 = |*_6 F_{5}|^2 \geq |(*_6 F_{5})|_{\bot}|^2$, allows us to prove that the Ramond-Ramond contributions to these lines are always negative (semi-)definite. We rewrite the final equation \eqref{FINALEr} as
\begin{empheq}[innerbox=\fbox, left=\!\!\!\!\!]{align}
& 2 e^{-2A} \tilde{{\cal R}}_4  = - \left|*_{\bot}H|_{\bot} + \varepsilon_p e^{\phi} F_{k-2}|_{\bot} \right|^2  - 2 e^{2\phi}\left| g_s^{-1} \tilde{*}_{\bot} \d e^{-4A} - \varepsilon_p  F_{k}^{(0)} \right|^2 \label{FINAL2Er} \\
& \phantom{2 e^{-2A} \tilde{{\cal R}}_4  =} - \sum_{a_{||}} \left| *_{\bot}( \d e^{a_{||}})|_{\bot} - \varepsilon_p e^{\phi} (\iota_{\del_{a_{||}}} F_k^{(1)} ) \right|^2\ - 2 {\cal R}_{||} - 2 {\cal R}_{||}^{\bot} + |H^{(2)}|^2 + 2|H^{(3)}|^2  \nn\\
& \phantom{2 e^{-2A} \tilde{{\cal R}}_4  =} - 2 e^{-2A} \left(\d \left(e^{8A} \tilde{*}_{\bot} \d e^{-4A} - e^{8A} \varepsilon_p g_s F_{k}^{(0)} \right)\right)_{\widetilde{\bot}} \nn\\
& \phantom{2 e^{-2A} \tilde{{\cal R}}_4  =} - e^{2\phi} \Bigg(  2 |F_{k+2}|^2  + (p-5) |F_{k+4}|^2 + \frac{1}{2} (|F_{5}|_{\bot}|^2 - |(*_6 F_{5})|_{\bot}|^2) \nn\\
& \phantom{2 e^{-2A} \tilde{{\cal R}}_4  = - e^{2\phi} \Bigg( } - \sum_{n\geq 2}^{p-3} (n-1) \left( |F_{k+2}^{(n)}|^2 + \frac{p-6}{2} |F_{k+4}^{(n)}|^2 + \frac{p-7}{4} |F_5^{(n)}|^2 \right) \Bigg) \nn \ ,
\end{empheq}
where the last two lines are a negative (semi-)definite contribution. The new combination \eqref{combifields} now appears. The integral version of this expression, \eqref{FINALint}, is similarly corrected. Turning to the no-go theorems, equations \eqref{cond0}, \eqref{require} and \eqref{nogo2456} still hold in view of \eqref{tracepargenEr}, the corrected version of \eqref{tracepargen}. They can however be refined with the new $H$-flux terms, towards
\begin{equation}
  2 {\cal R}_{6||} +  4 (\nabla\del \phi)_{6||} -  \frac{p-3}{2} \left({\cal R}_4  + 2 (\nabla\del \phi)_4 \right) - |H^{(2)}|^2 - 2|H^{(3)}|^2 \geq 0 \ ,
\end{equation}
for \eqref{cond0}. We deduce the following version of the main result, correcting \eqref{constraintcurv}
\begin{empheq}[innerbox=\fbox]{align}
& \mbox{There is no de Sitter vacuum for $p=4, 5$, or $6$, if the inequalities} \\
& - \int_{\widetilde{\mmm}} \widetilde{{\rm vol}}_{6}\, e^{2A}\,  \sum_{a_{||}} |(\d e^{a_{||}})|_{\bot}|^2\, < \, \int_{\widetilde{\mmm}} \widetilde{{\rm vol}}_{6}\, e^{2A} \left( 2 {\cal R}_{||} + 2 {\cal R}_{||}^{\bot} - |H^{(2)}|^2 - 2|H^{(3)}|^2 \right)\, < 0  \nn\\
& \mbox{are {\it not} satisfied.} \nn
\end{empheq}

\newpage

\pagenumbering{arabic}
\setcounter{page}{2}

\numberwithin{equation}{section}

\noindent\rule[1ex]{\textwidth}{1pt}
\vspace{-0.8cm}

\tableofcontents

\vspace{0.5cm}
\noindent\rule[1ex]{\textwidth}{1pt}
\vspace{0.4cm}

\section{Introduction}

Recent high precision cosmological observations \cite{Ade:2015tva, Ade:2015xua, Ade:2015ava, Ade:2015lrj} have set important constraints on models describing the early universe. The coming measures (see e.g.~\cite{Chen:2016vvw}) could even put a quantum gravity theory such as string theory under pressure \cite{Parameswaran:2016fqr}. It is thus a timely moment to address crucial pending questions of string cosmology, among which finding a metastable de Sitter vacuum. Our present accelerating universe is well described as a four-dimensional de Sitter space-time, and without a proposal for an evolution mechanism, this shape should remain when going back to the early times. For instance, the end-point or vacuum of inflation scenarios, where reheating occurs, is commonly considered to be a de Sitter vacuum (see also \cite{Conlon:2006tq}). In supergravity, many inflationary models have been proposed recently and been compared to the new experimental data, but few of them are realised completely within string theory (see e.g.~\cite{Andriot:2015aza}). This prevents from connecting them to U.V.~and quantum gravity aspects. To achieve this, one should be able to embed these scenarios in a string compactification, which requires to know the de Sitter vacuum, the internal compact geometry, etc. in full detail. In addition, this would allow to verify that all aspects of the compactification, e.g.~moduli stabilisation, are under control and do not spoil the inflation mechanism. With these motivations in mind, in the present paper we focus on the question of finding de Sitter vacua.

Several ideas have been proposed for how to construct de Sitter within string theory. A problem with these proposals is often the use of features that lack a full understanding in ten dimensions. A famous idea to achieve de Sitter is given in \cite{Kachru:2003aw} where anti-branes are used to uplift the value of the cosmological constant. Attempts to construct the underlying, backreacted, ten-dimensional solution supporting this scenario have encountered several challenges, starting with \cite{Bena:2009xk}. While this has been an active subject for years, the final outcome of \cite{Kachru:2003aw} remains unclear, see e.g. \cite{Bena:2016fqp}. Other remarks can be found e.g.~in \cite{deAlwis:2016cty, CaboBizet:2016qsa}.

While further proposals have been made to obtain a positive cosmological constant at the four-dimensional level (in particular by the use of non-geometric fluxes), we prefer here to remain in the simpler and somewhat safer (in terms of control on the compactification) setting of ten-dimensional classical de Sitter vacua. We consider standard ten-dimensional type II supergravities without $\alpha'$ corrections, supplemented by the Ramond-Ramond (RR) sources $D_p$-branes and orientifold $O_p$-planes; no Neveu-Schwarz source such as $\NS_5$-branes or Kaluza-Klein ($\KK$) monopoles are included. Relevant to us are vacua where the space-time is the warped product of a four-dimensional de Sitter space-time and a six-dimensional compact internal manifold $\mmm$: with a controlled value of the dilaton, this would be a valid classical background of string theory.\footnote{With some abuse of common terminology, a {\em vacuum} refers here to a solution of the equations of motion and Bianchi identities.} This ten-dimensional setting could be the only one where a classical de Sitter string background exists: indeed, such vacua have been ruled-out recently among supersymmetric heterotic string backgrounds \cite{Green:2011cn, Gautason:2012tb, Kutasov:2015eba, Quigley:2015jia}.

In our context, no-go theorems have also been established. To start with, standard ones for classical de Sitter vacua with compact internal geometries \cite{Gibbons:1984kp, deWit:1986xg, Maldacena:2000mw, Townsend:2003qv} are circumvented by requiring orientifolds. This is however far from being enough, and many refined no-go theorems have been worked-out \cite{Hertzberg:2007wc, Haque:2008jz, Caviezel:2008tf, Flauger:2008ad, Danielsson:2009ff, deCarlos:2009fq, Caviezel:2009tu, Wrase:2010ew, Burgess:2011rv, VanRiet:2011yc, Gautason:2013zw, Kallosh:2014oja}, most of them studying the corresponding four-dimensional scalar potential inspired by \cite{Silverstein:2007ac}, sometimes considering as well constraints on the slow-roll parameter for inflation or on the vacuum metastability. Note that the four-dimensional approach always has the drawback of considering smeared sources, and thus neglecting (or averaging) their backreaction (see e.g.~\cite{Blaback:2010sj, Junghans:2013xza} on this topic); the effectiveness of these models is also often debatable. In the present paper, we avoid such questions by working purely in ten dimensions and keeping the dependence on the warp factor and dilaton explicitly during our computation. From this whole literature, an outcome is that very few classical de Sitter vacua have been found, and none of them is metastable \cite{Caviezel:2008tf, Flauger:2008ad, Danielsson:2009ff, Caviezel:2009tu, Danielsson:2010bc, Danielsson:2011au}. Further work was dedicated directly to the stability problem \cite{Covi:2008ea, Shiu:2011zt, Danielsson:2012et, Junghans:2016uvg}, but no systematic explanation has been found for the tachyons appearing.

In this paper, we work in ten dimensions and focus on the existence of classical de Sitter vacua of type II supergravities with $D_p$ and $O_p$ sources, without ever considering the four-dimensional stability. We aim to provide general statements that would clarify the situation and refine the boundaries of the classical de Sitter landscape. To that end, we consider sources of one fixed dimension at a time, $3\leq p \leq 8$, which are also parallel, i.e.~not intersecting, or equivalently, having the same transverse subspace; see Section \ref{sec:compactif} for the detailed specifications on the sources and the internal geometry. In the particular case of a parallelizable internal manifold $\mmm$, having parallel sources would lead, after dimensional reduction, to a four-dimensional ${\cal N}=4$ gauged supergravity. There, to the best of our knowledge, only de Sitter solutions have been found with non-compact gaugings or gaugings with angles, which are unlikely to have a compactification origin. In addition, all known ten-dimensional (unstable) classical de Sitter vacua mentioned above have intersecting sources. A natural guess is then that a no-go theorem exists for parallel sources: the outcome of this work is very close to such a result. We first prove the following
\beq
\boxed{\mbox{There is no de Sitter vacuum for}\ p=3, \ 7,\ \mbox{or}\ 8.} \label{nogo378}
\eeq
The $p=3$ result was already derived in \cite{Blaback:2010sj}, whose methods act as an inspiration for the generalization to other $p$. Note that for $p=3$, sources are always parallel, making this result very general. In the other cases, we first reproduce in Section \ref{sec:O7O8} some results previously obtained in four dimensions. More importantly, we then derive the following
\begin{empheq}[innerbox=\fbox]{align}
& \mbox{There is no de Sitter vacuum for}\ p=4,\ 5,\ \mbox{or}\ 6, \label{nongo456intro}\\
& \mbox{if some curvature terms are}\ \leq -{\rm bound}\ {\rm or}\ \geq 0.  \nn
\end{empheq}
These curvature terms are related to curvatures of internal subspaces. As discussed in Section \ref{sec:nogo3456}, their value is constrained to a tight range, summarized in \eqref{constraintcurv}, leaving eventually very little room for de Sitter vacua, with parallel sources. These terms also vanish in many examples of Minkowski vacua. Finally, as a side result, we prove two more no-go theorems \eqref{nogodF0} and \eqref{nogocalib} in the smeared limit, building on the interesting expression \eqref{firstsquaregen}.

These results are derived thanks to appropriate combinations of ten-dimensional equations of motion and flux Bianchi identities, that isolate the unwarped four-dimensional curvature $\tilde{{\cal R}}_4$. For a de Sitter vacuum, we require the latter to be positive. On this aspect, the main result of the paper is the expression \eqref{FINAL} schematically given by
\beq
\boxed{ \tilde{{\cal R}}_4 = -\, \left(\mbox{BPS-like}\right)^2 \, -\, \left(\mbox{flux}\right)^2 \, -\, \mbox{curvature terms}\, +\, \mbox{total derivative} }  \label{FINALintro}
\eeq
It is inspired by the $p=3$ case of \cite{Blaback:2010sj} and generalizes \cite{Giddings:2001yu}. This expression makes the sign contributions to $\tilde{{\cal R}}_4$ apparent, and some of the above no-go theorems for $p=3,4,5,6$ are then easy to obtain; in particular, the curvature terms (and flux terms) vanish for $p=3$, leading to \eqref{finalO3}. For $p=7,8$, we followed \cite{Andriot:2015aza} to derive the appropriate expressions \eqref{4dfinalIIA} and \eqref{4dfinalIIB}. What is denoted ``BPS-like'' in \eqref{FINALintro} are interesting combinations: setting them to zero would fix the sourced RR flux $F_k$ (with $k=8-p$), and relate the flux $F_{k-2}$ to the $H$-flux, or at least components thereof. It generalizes the conditions obtained in \cite{Giddings:2001yu} for $p=3$, in particular the imaginary self-dual condition. This will be the topic of a companion paper \cite{Andriot:2016ufg}, where we focus on Minkowski vacua.

The paper is organised as follows. Conventions on ten-dimensional type II supergravities are given in the self-contained Appendix \ref{ap:conv}, and those are applied to our compactification setting as detailed in Section \ref{sec:compactif}. Then, we derive the no-go theorems for $p=7,8$ and further results for other $p$ values in Section \ref{sec:O7O8}. Different equation manipulations are then presented in Section \ref{sec:3456} to conclude and discuss the no-go theorems for $p=3,4,5,6$. We end with an outlook in Section \ref{sec:outlook}. Useful formulas and details of computations are given in Appendix \ref{ap:compute}. Appendix \ref{ap:calib} discusses extra conditions obtained by minimizing the energy of a $D_p$-brane.

\section{Compactification setting}\label{sec:compactif}

We consider ten-dimensional type IIA and IIB supergravities and use the conventions given in Appendix \ref{ap:conv}. We allow for Ramond-Ramond (RR) sources, namely $D_p$-branes and orientifold $O_p$-planes, but for no further ingredient. In particular, we do not include $\NS_5$-branes or $\KK$-monopoles, one problem with those being the Bianchi identity tadpole cancelation. In this section, we specify to a compactification setting and detail our notations. The ten-dimensional space-time is a warped product of a four-dimensional maximally symmetric space-time (anti-de Sitter, Minkowski, de Sitter) along directions $\d x^{\mu}$ and a six-dimensional (internal) compact manifold $\mmm$ along directions $\d y^m$. The metric is written accordingly
\beq
\d s^2= e^{2A(y)} \tilde{g}_{\mu\nu} (x) \d x^\mu \d x^\nu + g_{mn} (y) \d y^m \d y^n \ .\label{10dmetric}
\eeq
The warp factor is $e^A$. A tilde denotes quantities without the warp factor, i.e.~where it has been explicitly extracted; we also dub such quantities as ``smeared''. Looking for a vacuum, we will require to preserve Lorentz invariance in four dimensions. A first consequence is that the dilaton is restricted to depend only on internal coordinates. Further, the fluxes $F_0, F_1, F_2, F_3, H$ have to be purely internal (in components and coordinate dependence), and $F_4^{10}$ and $F_5^{10}$ can have four-dimensional components in a constrained manner. With the unwarped, and warped, four-dimensional volume form denoted $\widetilde{{\rm vol}}_4= \sqrt{|\tilde{g}_4|} \d^4 x$, and ${\rm vol}_4$, respectively, one can have
\beq
F_4^{10}= F_4^4 + F_4 \ {\rm with}\ F_4^4= \widetilde{{\rm vol}}_4\, f_4 \ , \ \ F_5^{10}= F_5^4 + F_5 \ {\rm with}\ F_5^4= \widetilde{{\rm vol}}_4 \w f_5 \ ,\label{fluxes}
\eeq
with an internal scalar $f_4$, and internal forms $F_4$, $F_5$, $f_5$. We introduce as a notation an internal $6$-form $F_6$ such that $f_4=e^{4A} *_6 F_6$; because $F_5^4= -*_{10} F_5$, one obtains $f_5=-e^{4A} *_6 F_5$, so
\beq
F_4^4= {\rm vol}_4 \w *_6 F_6 \ ,\ F_5^4= - {\rm vol}_4 \w *_6 F_5 \ .
\eeq
Since $|*_6 F_6|^2=|F_6|^2$, we deduce $|F_4^{10}|^2=|F_4|^2 - |F_6|^2$, and $|F_5|^2=-|F_5^4|^2= |*_6 F_5|^2 $.

We now impose few restrictions on the sources $D_p$ and $O_p$ and the related internal geometry. Here are first some properties of the sources:
\begin{enumerate}
  \item Because of four-dimensional Lorentz invariance, the sources have to be space-time filling, meaning that their world-volume spans the whole four-dimensional space-time, and possibly wraps some internal subspace; this restricts $p\geq 3$, and we consider $p\leq8$. \label{it:pt1}
  \item We consider for each source that $-\imath^*[b] + \mathcal{F} = 0$ (see Appendix \ref{ap:conv} for more details). We also consider them to be BPS, giving $\mu_p=T_p$. \label{it:pt2}
  \item We restrict ourselves to sources of only one fixed size $p$. \label{it:pt3}
  \end{enumerate}
Further, we need in this paper (for $p=4,5,6$) to formalise the common idea of the subspace of $\mmm$ wrapped by a source, and the one transverse to it. Let us start by presenting our formal characterization of these subspaces for one source, and then give examples of manifolds $\mmm$ captured by our description. Locally, one can always reach the orthonormal basis (in which we will work), where the unwarped (smeared) internal metric is written $\d \tilde{s}^2 = \delta_{ab} \tilde{e}^a \tilde{e}^b$, with the orthonormal basis one-forms $\tilde{e}^{a}=\tilde{e}^a{}_m \d y^m$. We restrict ourselves to a setting where the set $\{ \tilde{e}^a\}$ can be split globally into two sets, one of $(p-3)$ ``parallel'' one-forms $\{ \tilde{e}^{a_{||}} \}$ and one of $(9-p)$ ``transverse'' one-forms $\{ \tilde{e}^{a_{\bot}} \}$. This global requirement translates mathematically as the structure group of the cotangent (frame) bundle being reduced from O(6) to $O(p-3) \times O(9-p)$, or a subgroup thereof. This does not imply that each $\tilde{e}^{a_{||}}$ or $\tilde{e}^{a_{\bot}}$ is globally defined, only the whole sets are: for instance, parallel one-forms may still get exchanged among themselves, but not with transverse ones. We are thus not restricting to parallelizable manifolds; those, such as twisted tori, are only one example captured by our setting, since their cotangent bundle structure group is the identity. Another example is the direct product of two manifolds, not necessarily parallelizable, e.g.~the product of a two- and a four-sphere: the metric can then be taken to be block diagonal in some coordinate basis, and this block structure is globally preserved in the orthonormal basis. More generally, any fibered manifold, or fiber bundle (with metric), fits in our setting. In a fiber bundle, horizontal one-forms (on the base) are well-defined, but one needs a connection to define vertical ones; this is provided by the metric, which defines vertical one-forms as being orthogonal to horizontal ones. By definition, the metric is then block diagonal in this (horizontal/vertical) one-form basis, where one-forms are globally defined. A standard example is the fibration of a circle (along $y$) over a base ${\cal B}$, where the metric is given by $\d s_{{\cal B}}^2 + g_{{\cal F}} (\d y + A)^2$, with a base one-form $A$ that makes $\d y + A$ globally defined in $\mmm$. Natural generalizations of this formula exist for principle fiber bundles. So for a fibered manifold, the metric is globally block diagonal in some one-form basis, and this structure can be brought to the orthonormal basis, allowing to define globally the two sets $\{ \tilde{e}^{a_{||}} \}$ and $\{ \tilde{e}^{a_{\bot}} \}$; a fibered manifold is thus captured by our setting. While the physically relevant six-dimensional internal manifold is the one just discussed, namely the underlying unwarped or smeared one, we now deform it by introducing the warp factor, with $e^a=e^{\pm A} \tilde{e}^{a}$. More precisely, for each source, a 10d metric can now be written from \eqref{10dmetric}, as
\beq
\d s^2_{10}= e^{2A} ( \d \tilde{s}^2_{4} + \d \tilde{s}^2_{6||} ) + e^{-2A} \d \tilde{s}^2_{6\bot} \ ,\ {\rm where}\ \d \tilde{s}^2_{6||}= \delta_{ab} \tilde{e}^{a_{||}} \tilde{e}^{b_{||}} \ ,\ \d \tilde{s}^2_{6\bot}= \delta_{ab} \tilde{e}^{a_{\bot}} \tilde{e}^{b_{\bot}} \ .\label{metricwarp}
\eeq
The warp factor is restricted to depend only on transverse directions, in the following sense: $\forall\, {}_{a_{||}},\ e^m{}_{a_{||}} \del_m A = \del_{a_{||}} A = e^{-A} \del_{\tilde{a}_{||}} A = 0$. With some abuse, we will call collectively the sets of parallel or transverse directions as the parallel or transverse subspaces, and define naturally their ``volume forms'': ${\rm vol}_{||}= \epsilon_{a_{1||} \dots a_{(p-3)||}} e^{a_{1||}} \w \dots \w e^{a_{(p-3)||}}$, and similarly for ${\rm vol}_{\bot}$. These volume forms are globally defined by construction. With our ordering conventions, one has
\bea
& {\rm vol}_{4} \w {\rm vol}_{||} \w {\rm vol}_{\bot} = {\rm vol}_{10} = \d^{10} x \sqrt{|g_{10}|} \ ,\label{volrel}\\
& {\rm vol}_{||} \w {\rm vol}_{\bot} = {\rm vol}_{6} = \d^{6} y \sqrt{|g_{6}|} \ ,\ *_6 {\rm vol}_{\bot} = (-1)^{9-p} {\rm vol}_{||} \ , \ *_6 {\rm vol}_{||} = {\rm vol}_{\bot} \ .\nn
\eea
Let us summarize the above geometric considerations and relate this to the sources:
\begin{enumerate}
\setcounter{enumi}{3}
  \item \label{it:pt4} For each source, we assume the existence of a global split of $\{ \tilde{e}^a\}$ into $\{ \tilde{e}^{a_{||}} \}$ and $\{ \tilde{e}^{a_{\bot}} \}$, i.e.~we require the structure group of the cotangent bundle to be a subgroup of $O(p-3) \times O(9-p)$; this includes e.g.~fibered manifolds. The 10d metric is then given by \eqref{metricwarp}, and one defines accordingly the (global) volume forms ${\rm vol}_{||}$ and ${\rm vol}_{\bot}$ satisfying the above properties. Finally, each source is considered to wrap its parallel subspace, meaning that its world-volume form is given by
\beq
\d^{p+1} \xi \sqrt{|\imath^*[g_{10}]|}= \imath^*[{\rm vol}_{4} \w {\rm vol}_{||}] \ .\label{worldvolform}
\eeq
\end{enumerate}
We finally specify two more restrictions on the sources and related geometry.
\begin{enumerate}
\setcounter{enumi}{4}
  \item We consider all sources to be parallel, meaning having the same transverse directions. Note that for $p=3$, this is not an assumption. As a consequence, the metric \eqref{metricwarp} specified for each source holds in general, using only one warp factor. We also recall that the coordinate dependence of $\d \tilde{s}^2_{6||},\ \d \tilde{s}^2_{6\bot}$ has so far been left generic. \label{it:pt5}
  \item Even though most of our computations are done locally (with equations of motion), we will need in the very end to perform an integral. We will then require the transverse unwarped subspace to be a compact manifold without boundary.\footnote{The (naive) singularities at the sources loci, and the cut-off at the string scale required to stay in a valid supergravity regime, may raise some doubt on the use of this ``no boundary'' assumption, especially when getting close to the sources (see also \cite{Burgess:2011rv}). One should first note that this assumption is made on an unwarped space, described by $\tilde{g}_{mn}$, which intrinsically does not have any source-related singularity: if any, those are present in the warp factor, outside of $\tilde{g}_{mn}$. Secondly, when this assumption is used here to integrate a total derivative, the integrand is not a gauge potential (as e.g. in \cite{Burgess:2011rv}) but a flux and derivatives of warp factor. This integration is very close to the standard one of the Laplacian of the warp factor in the flux Bianchi identity, which always gives zero. We thus believe that the presence of sources does not affect here our use of this assumption.} If for instance $\mmm$ is a fibered manifold and the transverse subspace is the base, one can easily find examples where this property holds. \label{it:pt6}
\end{enumerate}

Thanks to these properties, the Bianchi identities (BI) for $F_4^{10}$ and $F_5^{10}$ impose $f_4$ and $f_5$ to be closed; then, the RR BI can be restricted to the internal forms $F_k$ only, towards
\bea
&\d F_k - H \w F_{k-2} = -\varepsilon_p\, 2 \kappa_{10}^2\,  T_p \!\!\! \sum_{p-{\rm sources}} \!\!\! c_p\, \delta^{\bot}_{9-p} =  \varepsilon_p\, \frac{T_{10}}{p+1} {\rm vol}_{\bot} \ , \label{BI2}\\
& \mbox{for}\ 0 \leq k=8-p \leq 5\ ,\ \varepsilon_p=(-1)^{p+1} (-1)^{\left[\frac{9-p}{2} \right]} \ ,\nn
\eea
where $F_{-1} =F_{-2}=0$, and one uses \eqref{T10} for $T_{10}$. Given the right-hand side of \eqref{BI2}, we will need to project forms on the transverse directions. To that end, we introduce the following notations: for a form $G$, we denote its projection on the transverse directions with $G|_{\bot}$ or $(G)|_{\bot}$, i.e.~the form obtained by keeping only its components entirely along those directions. In addition, if $G$ is a $(9-p)$-form, $(G)_{\bot}$ denotes the coefficient of this form on the transverse world-volume, i.e.~$G|_{\bot}= (G)_{\bot} {\rm vol}_{\bot}$; one has equivalently $(G)_{\bot} = *_{\bot} G|_{\bot}$. We deduce that the BI \eqref{BI2} gives after projection
\beq
(\d F_k)_{\bot} - (H \w F_{k-2})_{\bot} =  \varepsilon_p\, \frac{T_{10}}{p+1} \ . \label{BI3}\\
\eeq
Note that $(H \w F_{k-2})|_{\bot}= H|_{\bot} \w F_{k-2}|_{\bot}$. Using that $A\w * B=B\w *A$ for forms $A$ and $B$ of same degree, we can show that $*_{\bot} H|_{\bot} \w *_{\bot} F_{k-2}|_{\bot}= F_{k-2}|_{\bot} \w *_{\bot}^2 H|_{\bot}= H|_{\bot} \w F_{k-2}|_{\bot}$. From this we conclude, for any sign $\varepsilon$
\beq
\left|*_{\bot} H|_{\bot} + \varepsilon e^{\phi}  F_{k-2}|_{\bot} \right|^2 = |H|_{\bot}|^2 + e^{2\phi} |F_{k-2}|_{\bot}|^2 + 2 \varepsilon e^{\phi} (H\w F_{k-2})_{\bot} \ , \label{blasquare}
\eeq
where the definition of the square is given below \eqref{Hodge}. This formula and reasoning will be useful. For completeness, we give the fluxes' equations of motion (e.o.m.) expressed in terms of internal quantities, considering no source contribution to the $b$-field e.o.m.
\bea
& e^{-4A} \d(e^{4A} *_6 F_q ) + H \w *_6 F_{q+2} = 0\ \ (1 \leq q \leq 4) \ ,\\
& e^{-4A} \d (e^{4A-2\phi} *_6 H) -  \sum_{0\leq q \leq 4} F_{q} \w *_6 F_{q+2} = 0 \ .
\eea

We turn to the dilaton e.o.m. and Einstein equation. We denote ${\cal R}_{10}= g^{MN} {\cal R}_{MN}$, and
\beq
{\cal R}_4= g^{MN} {\cal R}_{MN=\mu\nu} \ ,\ {\cal R}_6= g^{MN} {\cal R}_{MN=mn}={\cal R}_{10} - {\cal R}_{4} \ , \ (\nabla\del \phi)_4= g^{MN=\mu\nu}\nabla_{M}\del_{N} \phi \ .
\eeq
The dilaton e.o.m., the ten-dimensional Einstein trace, and the four-dimensional one,\footnote{Let us detail the indices counting for $F_5^{10}$: the four-dimensional trace selects the $F_5^4$ piece giving
\beq
\frac{g^{\mu\nu}}{2\cdot 4!}  F_{5\ \mu PQRS}^{4}F_{5\ \nu}^{4 \ \ PQRS} = \frac{ g^{\mu\nu}}{2\cdot 3!} F_{5\ \mu \pi\rho \tau s}^{4}F_{5\ \nu}^{4 \ \ \pi\rho \tau s} = \frac{2}{4!} F_{5\ \mu \pi\rho \tau s}^{4}F_{5}^{4 \ \mu \pi\rho \tau s} = \frac{2}{5\cdot 4!}  F_{5\ MPQRS}^{4}F_{5}^{4 \ MPQRS} = 2 |F_5^4|^2 \ .\nn
\eeq} are
\bea
& \hspace{-0.1in} 2 {\cal R}_{10} + e^{\phi} \frac{T_{10}}{p+1} -|H|^2 + 8(\Delta \phi - |\del \phi|^2 ) = 0 \ ,\label{dileom2}\\
& \hspace{-0.1in} 4 {\cal R}_{10}  + \frac{e^{\phi}}{2} {T}_{10} - |H|^2 - \frac{e^{2\phi}}{2} \sum_{q=0}^6 (5-q) |F_q|^2 -20 |\del \phi|^2 + 18 \Delta \phi = 0\ , \label{10dtrace}\\
& \hspace{-0.1in} {\cal R}_4 - 2{\cal R}_{10} - \frac{2 e^{\phi}}{p+1} {T}_{10} + |H|^2 + e^{2\phi} \sum_{q=0}^6 |F_q|^2  +2 (\nabla\del \phi)_4 + 8 |\del \phi|^2 - 8 \Delta \phi = 0 \ ,\label{4dtrace}
\eea
where one should only consider even/odd RR fluxes in IIA/IIB, and we used the above properties (we only used Point \ref{it:pt1} through \ref{it:pt4} for the sources), giving in particular $g^{MN} T_{MN=\mu\nu} = 4 T_{10} /(p+1)$. These scalar equations will be combined to express ${\cal R}_4$ in terms of a limited number of ingredients.

\section{No de Sitter vacuum for $O_7$, $O_8$, and more no-go theorems}\label{sec:O7O8}

Given the context presented in Section \ref{sec:compactif}, we prove here that there cannot be any de Sitter vacuum for $p=7,8$ sources, and get constraints for the other $p$, that can be viewed as no-go theorems. We derive these results in ten dimensions; for $p=7,8$, this is done {\it without smearing}. This reproduces known results obtained in \cite{Danielsson:2009ff, Wrase:2010ew} from a four-dimensional approach, that uses conditions for a vacuum but also for its stability.

We proceed as in \cite{Andriot:2015aza}: we first use the dilaton e.o.m.~to eliminate ${T}_{10}$ in respectively the ten- and four-dimensional traces; we get (with even/odd RR fluxes in IIA/IIB)
\bea
&\hspace{-0.1in} (p-3)\left( -2  {\cal R}_{10}  + |H|^2 + 8 |\del \phi|^2 - 8 \Delta \phi \right) + 2 |H|^2 - e^{2\phi} \sum_{q=0}^6 (5-q) |F_q|^2 -2 e^{2\phi} \Delta e^{-2\phi} = 0  \label{10dtracesansT10}\\
&\hspace{-0.1in} 3 {\cal R}_4 = - 2{\cal R}_{6} + |H|^2 - e^{2\phi} \sum_{q=0}^6 |F_q|^2 -2 (\nabla\del \phi)_4 + 8 |\del \phi|^2 - 8 \Delta \phi \ .\label{4dtracesansT10}
\eea
with $-2 |\del \phi|^2 + \Delta \phi= -\frac{1}{2} e^{2\phi} \Delta e^{-2\phi}$. We now multiply \eqref{4dtracesansT10} by $(p-3)$, insert \eqref{10dtracesansT10} and get
\beq
(p-3) {\cal R}_4  = - 2 |H|^2 + e^{2\phi} \sum_{q=0}^6 (8-q-p) |F_q|^2 + 2 e^{2\phi} \Delta e^{-2\phi} -2 (p-3)(\nabla\del \phi)_4 \ .\label{4dtracefinal}
\eeq
Now, the warp factor and dilaton terms need to be computed: this is done in Appendix \ref{ap:compute} using the metric \eqref{10dmetric}. As mentioned there, we pick in this paper the following standard dilaton value, that provides natural simplifications
\beq
e^{\phi}= g_s e^{A(p-3)} \label{dilatonwarp} \ ,
\eeq
where $g_s$ is a constant. This value might be derived for $p=7, 8$ from \eqref{4dtracefinal}, but we simply impose it here for all $p$. Note that this prevents us from capturing the non-perturbative F-theory solutions. As shown with \eqref{qttyap}, this value allows to obtain
\beq
(p-3) {\cal R}_4 - 2 e^{2\phi} \Delta e^{-2\phi} + 2 (p-3) (\nabla\del \phi)_4 = (p-3) e^{-2A} \tilde{{\cal R}}_4 \ ,
\eeq
where $\tilde{{\cal R}}_4$ is the four-dimensional Ricci scalar built from $\tilde{g}_{\mu\nu}$. We conclude in IIA and IIB
\begin{empheq}[innerbox=\fbox, left=\!\!\!\!]{align}
\frac{(p-3)}{e^{2A}} \tilde{{\cal R}}_4  = & - 2 |H|^2 + e^{2\phi} \left( (8-p) |F_0|^2 + (6-p) |F_2|^2 + (4-p) |F_4|^2 + (2-p) |F_6|^2 \right), \label{4dfinalIIA} \\
\frac{(p-3)}{e^{2A}} \tilde{{\cal R}}_4  = & - 2 |H|^2 + e^{2\phi} \left( (7-p) |F_1|^2 + (5-p) |F_3|^2 + (3-p) |F_5|^2 \right). \label{4dfinalIIB}
\end{empheq}
These equations have an interesting interpretation for $p\neq 3$: if the $D_p$ and $O_p$ source magnetically the flux $F_k$, the coefficient in front of $F_k$ precisely vanishes \cite{Andriot:2015aza}; $\tilde{{\cal R}}_4$ is then only given by the non-sourced fluxes.

We now study the possibility of getting a de Sitter vacuum, i.e.~$\tilde{{\cal R}}_4>0$. From \eqref{4dfinalIIA} and \eqref{4dfinalIIB}, the result is clear for $p=7,8$:
\beq
\boxed{\mbox{There is no de Sitter vacuum for}\ p=7\ \mbox{or}\ p=8.} \label{nogo78}
\eeq
Let us make a comment: we only used combinations of e.o.m. which required Points \ref{it:pt1}, \ref{it:pt2} and \ref{it:pt3}, from Section \ref{sec:compactif}, on the sources. In particular, we did not require Point \ref{it:pt5} on the assumption of parallel sources. So this result on $p=7,8$ could be extended to intersecting sources.

We now turn to the sources with $3 \leq p \leq 6$ in the smeared limit, in which the dilaton and warp factor are taken constant. We denote collectively $(\phi)$ the dilaton terms to be neglected.
\begin{itemize}
  \item $p=6$: equating \eqref{4dtracefinal} with \eqref{4dtracesansT10}, we get (as in \cite{Andriot:2010ju})
      \bea
      \frac{9}{2} {\cal R}_4 &= 3 \left( e^{2\phi} \left( |F_0|^2 - |F_4|^2 -2 |F_6|^2 \right) - |H|^2 \right) + (\phi) \label{R4O6}\\
       &= - 2 {\cal R}_6 - e^{2\phi} \left( |F_2|^2 + 2 |F_4|^2 + 3 |F_6|^2 \right)  + (\phi) \ . \nn
      \eea
  For de Sitter, one needs $F_0\neq 0$ and ${\cal R}_6 <0$ of sufficient magnitude to overtake the remaining possible non-zero terms, as pointed-out already in \cite{Haque:2008jz}.
  \item $p=5$: equating three halves of \eqref{4dtracefinal} with \eqref{4dtracesansT10}, we get
      \beq
      4 {\cal R}_4 = 4 \left(e^{2\phi} (|F_1|^2 - |F_5|^2 ) - |H|^2 \right) + (\phi) = - 2 {\cal R}_6 - e^{2\phi} (|F_3|^2 + 2 |F_5|^2 ) + (\phi) \ . \label{R4O5}
      \eeq
  For de Sitter, one needs $F_1\neq 0$ and ${\cal R}_6 <0$, of sufficient magnitude.
  \item $p=4$: equating three times \eqref{4dtracefinal} with \eqref{4dtracesansT10}, we get (as in \cite{Andriot:2015aza})
      \bea
      \frac{7}{2} {\cal R}_4 &= 7 \left(e^{2\phi} (2 |F_0|^2 + |F_2|^2- |F_6|^2 ) - |H|^2 \right) + (\phi) \label{R4O4} \\
      &= - 2 {\cal R}_6 + e^{2\phi} (|F_0|^2 - |F_4|^2 -2 |F_6|^2 ) + (\phi) \ . \nn
      \eea
  For de Sitter, one needs $F_0\neq0$, or $F_2\neq0$ and ${\cal R}_6 <0$, all of sufficient magnitude.
  \item $p=3$: \eqref{4dtracefinal} and \eqref{4dtracesansT10} give (using \eqref{dilatonwarp} for the dilaton)
      \beq
      3 {\cal R}_4 = - 2 {\cal R}_6 + e^{2\phi} (|F_1|^2 - |F_5|^2 )  \ ,\ \  2 e^{2\phi} |F_1|^2 = |H|^2 - e^{2\phi} |F_3|^2 \ .\label{R4O3bis}
      \eeq
  For de Sitter, one needs ${\cal R}_6 <0$, or $F_1\neq 0$ and $H\neq 0$, all of sufficient magnitude.
\end{itemize}
These are limited results, valid in the smeared limit. In the next section we will make use of the BI which will allow us to put further restrictions on the possibility of de Sitter vacua.

\section{No de Sitter vacuum for $O_3$, no-go theorems for $O_4$, $O_5$, $O_6$}\label{sec:3456}

\subsection{First manipulations}

In Section \ref{sec:O7O8}, we combined the e.o.m.~to eliminate $T_{10}$. Here we will eliminate ${\cal R}_{10}$ (or ${\cal R}_{6}$), and make a further step by using the BI for $T_{10}$. Finally, we will use another equation, the trace of the Einstein equation along the internal parallel directions, to rewrite the result more conveniently: this will bring us to the no-go theorems.

We start by combining the dilaton e.o.m.~and the four-dimensional trace to get
\beq
{\cal R}_4 = e^{\phi} \frac{{T}_{10}}{p+1} - e^{2\phi} \sum_{q=0}^6 |F_q|^2 - 2 (\nabla\del \phi)_4 \ , \label{R4T10F}
\eeq
with even/odd RR fluxes in IIA/IIB. Note that in smeared limit where the dilaton and warp factor are constant, one concludes that de Sitter needs ${T}_{10} > 0$ \cite{Maldacena:2000mw}; this requirement not only means having $O_p$, but also that they contribute more than $D_p$. We now combine the dilaton e.o.m.~with the ten-dimensional trace and get
\beq
(p-3) e^{\phi} \frac{{T}_{10}}{p+1}  +2 |H|^2 - e^{2\phi} \sum_{q=0}^6 (5-q) |F_q|^2 -8 |\del \phi|^2 + 4 \Delta \phi = 0\ . \label{10dtracesansR10}
\eeq
Equation \eqref{R4T10F} is multiplied by $-(p+1)$, and added to \eqref{10dtracesansR10}, giving
\beq
{\cal R}_4 + 2 (\nabla\del \phi)_4 = -\frac{1}{p+1} \bigg( -8 |\del \phi|^2 + 4 \Delta \phi -4 e^{\phi} \frac{{T}_{10}}{p+1} +2|H|^2 + e^{2\phi} \sum_{q=0}^6 (p+q-4) |F_q|^2  \bigg) \ . \label{R4T10HF0}
\eeq
From now on, we use notations of \eqref{BI2}, where the magnetically sourced flux is $F_k$ with $0 \leq k=8-p \leq 5$, and $F_{-1}=F_{-2}=F_7=F_8=F_9=F_{10}=F_{11}=0$. Then, \eqref{R4T10HF0} gets rewritten as
\bea
{\cal R}_4 + 2 (\nabla\del \phi)_4 = -\frac{2}{p+1} \bigg(& -4 |\del \phi|^2 + 2 \Delta \phi -2 e^{\phi} \frac{{T}_{10}}{p+1} +|H|^2 \label{R4T10HF}\\
& + e^{2\phi} (  |F_{k-2}|^2 + 2 |F_{k}|^2 + 3 |F_{k+2}|^2 + 4 |F_{k+4}|^2 + 5 |F_{k+6}|^2 )  \bigg)  \ . \nn
\eea
We now use the BI projected on transverse directions \eqref{BI3} to replace $T_{10}$. With \eqref{blasquare}, we get
\begin{empheq}[innerbox=\fbox, left=\!\!\!\!]{align}
{\cal R}_4 + 2 (\nabla\del \phi)_4  = -\frac{2}{p+1} \bigg( & -4 |\del \phi|^2 + 2 \Delta \phi -  2 \varepsilon_p e^{\phi} (\d F_k)_{\bot}  + \left|*_{\bot}H|_{\bot} + \varepsilon_p e^{\phi} F_{k-2}|_{\bot} \right|^2  \label{firstsquaregen}\\
& +|H|^2 - |H|_{\bot}|^2 + e^{2\phi} (  |F_{k-2}|^2 - |F_{k-2}|_{\bot}|^2 ) \nn\\
& + e^{2\phi} ( 2 |F_{k}|^2 + 3 |F_{k+2}|^2 + 4 |F_{k+4}|^2 + 5 |F_{k+6}|^2 )  \bigg)  \ . \nn
\end{empheq}
Let us make a few comments. In the smeared limit, the only term in the right-hand side with indefinite sign is $(\d F_k)_{\bot}$. There are thus two interesting subcases to mention. First, we get
\beq
\boxed{\mbox{There is no (smeared) de Sitter vacuum if in the smeared limit}\ (\d F_k)_{\bot} \rightarrow 0\ .}\label{nogodF0}
\eeq
For instance, in the Minkowski vacua of \cite{Giddings:2001yu, Andriot:2015sia}, $F_k$ is only given by a $\del A$ which vanishes in the smeared limit. Deformations of the vacuum preserving this property will then not give de Sitter. Second, as derived in Appendix \ref{ap:calib}, Minkowski vacua with calibrated sources satisfy
\beq
F_{k} = (-1)^p \varepsilon_p e^{-4A} *_6 \d\left( e^{4A -\phi} {\rm vol}_{||} \right) \ . \label{Fkcalib}
\eeq
This calibration condition, related to the source energy minimization, is automatically satisfied in Minkowski supersymmetric vacua \cite{Martucci:2005ht}. From \eqref{Fkcalib}, one can show
\beq
\int_{\mmm} 2e^{\phi} \varepsilon_p f\, (\d F_k )_{\bot} {\rm vol}_{6} = \int_{\mmm}  2e^{2\phi}f\, |F_k|^2\, {\rm vol}_{6} \ ,  \label{calibint}
\eeq
with $f= e^{4A-2\phi}$. Upon integration, the $(\d F_k )_{\bot}$ term in \eqref{firstsquaregen} is then compensated by the $|F_k|^2$ one, which leads us to conclude on de Sitter in the smeared limit
\begin{empheq}[innerbox=\fbox]{align}
& \mbox{There is no (smeared) de Sitter vacuum if sources are Minkowski-calibrated,} \label{nogocalib}\\
& \mbox{i.e.~if (\ref{Fkcalib}) holds.}\nn
\end{empheq}
For instance, deforming a supersymmetric Minkowski vacuum while preserving \eqref{Fkcalib}, by e.g.~adding more fluxes or changing part of the geometry, will not give de Sitter.

To go further, we need to characterise $(\d F_k )_{\bot}$. To that end, we use the orthonormal basis: the internal metric is expressed with vielbeins as $g_{mn}= e^a{}_m e^b{}_n \delta_{ab}$ and we denote $\del_a= e^m{}_a \del_m,\ e^a=e^a{}_m \d y^m$. In the following we will refer to the Latin indices starting with $a$ as \emph{flat indices}. The ``geometric flux'' $f^{a}{}_{bc}$ is defined as
\beq
\d e^a= -\frac{1}{2} f^{a}{}_{bc} e^b\w e^c \ \Leftrightarrow \ f^{a}{}_{bc} = 2 e^a{}_m \del_{[b} e^m{}_{c]} = - 2 e^m{}_{[c} \del_{b]} e^a{}_{m} \label{fabc} \ .
\eeq
In Section \ref{sec:compactif}, we further introduced the parallel and transverse flat indices, together with the metric \eqref{metricwarp}. We thus decompose $F_k$ on its parallel or transverse (flat) components
\beq
F_k= \frac{1}{k!} F^{(0)}_{k\ a_{1\bot}\dots a_{k\bot}} e^{a_{1\bot}} \w \dots \w e^{a_{k\bot}} + \frac{1}{(k-1)!} F^{(1)}_{k\ a_{1||}\dots a_{k\bot}} e^{a_{1||}} \w e^{a_{2\bot}} \w \dots \w e^{a_{k\bot}} + \dots \ , \label{decompoflux}
\eeq
where terms with at least two parallel directions have been left out. By definition, $F_k^{(0)}=F_k|_{\bot}$; we also take for convenience $F_0=F_0|_{\bot}$ and $F_0^{(1)}=0$. One deduces
\beq
(\d F_k)|_{\bot}  = (\d F_k^{(0)})|_{\bot} + (\d F_k^{(1)})|_{\bot}\ , \ \ (\d F_k^{(1)})|_{\bot} =  (\iota_{\del_{a_{||}}} F_k^{(1)}) \w (\d e^{a_{||}})|_{\bot} \ ,\label{dFgen}
\eeq
with $\iota_{V}$ the contraction by a vector $V$, e.g.~$\iota_{\del_{a_{||}}} e^{b_{||}} = \delta_{a_{||}}^{b_{||}}$, and $(\d e^{a_{||}})|_{\bot} = -\frac{1}{2} f^{a_{||}}{}_{b_{\bot}c_{\bot}} e^{b_{\bot}}\w e^{c_{\bot}}$. Proceeding similarly to \eqref{blasquare}, we further have
\bea
& \sum_{a_{||}} \left| *_{\bot}( \d e^{a_{||}})|_{\bot} - \varepsilon_p e^{\phi} (\iota_{\del_{a_{||}}} F_k^{(1)} ) \right|^2 = \sum_{a_{||}} e^{2\phi} |(\iota_{\del_{a_{||}}} F_k^{(1)})|^2 + \sum_{a_{||}} |(\d e^{a_{||}})|_{\bot}|^2 \label{squareF1}\\
& \phantom{\sum_{a_{||}} \left| *_{\bot}( \d e^{a_{||}})|_{\bot} - \varepsilon_p e^{\phi} (\iota_{\del_{a_{||}}} F_k^{(1)} ) \right|^2 =} -  2 \varepsilon_p e^{\phi} ((\iota_{\del_{a_{||}}} F_k^{(1)}) \w (\d e^{a_{||}})|_{\bot})_{\bot} \nn\\
& \mbox{with}\ \sum_{a_{||}} e^{2\phi} |(\iota_{\del_{a_{||}}} F_k^{(1)})|^2 = e^{2\phi} |F_k^{(1)}|^2 \ ,\quad \sum_{a_{||}} |(\d e^{a_{||}})|_{\bot}|^2 = \frac{1}{2} \delta^{be} \delta^{cf} \delta_{ad} f^{a_{||}}{}_{b_{\bot}c_{\bot}} f^{d_{||}}{}_{e_{\bot}f_{\bot}} \ .\nn
\eea
We thus reconstruct interesting squares from $(\d F_k )_{\bot}$ at the cost of introducing the geometric contributions $|(\d e^{a_{||}})|_{\bot}|^2$. Those actually appear in curvature terms, present in the trace of the Einstein equation \eqref{EinstIIA} or \eqref{EinstIIB} along internal parallel flat directions. So we turn to this trace, and denote ${\cal R}_{6||}$ the trace of the ten-dimensional Ricci tensor along internal parallel flat directions. We obtain
\bea
{\cal R}_{6||} + 2 (\nabla\del \phi)_{6||} &= \frac{p-3}{4} \left({\cal R}_4 +  2 (\nabla\del \phi)_4 + 2e^{2\phi} |F_{6}|^2  \right) \label{traceparIIA}\\
& + \frac{1}{2} \left(|H|^2 - |H|_{\bot}|^2 + e^{2\phi} (  |F_{2}|^2 - |F_{2}|_{\bot}|^2 + |F_{4}|^2 - |F_{4}|_{\bot}|^2 \right) \nn \\
{\cal R}_{6||} + 2 (\nabla\del \phi)_{6||} &= \frac{p-3}{4} \left({\cal R}_4 +  2 (\nabla\del \phi)_4 + e^{2\phi} |F_{5}|^2  \right) \label{traceparIIB}\\
& + \frac{1}{2} \left(|H|^2 - |H|_{\bot}|^2 + e^{2\phi} (  |F_{1}|^2 - |F_{1}|_{\bot}|^2 + |F_{3}|^2 - |F_{3}|_{\bot}|^2 \right) \nn\\
& + \frac{1}{4} e^{2\phi} \left(  |F_{5}|^2 - |F_{5}|_{\bot}|^2 - |*_6 F_{5}|^2 + |(*_6 F_{5})|_{\bot}|^2 \right)  \ ,\nn
\eea
where we used the four-dimensional trace of the Einstein equation for terms in $p-3$. The above is valid for $0 \leq k=8-p \leq 5$; for $p=3$ where all internal directions are transverse, we take as a definition of the left-hand side that it vanishes. A generic rewriting of the above is
\bea
\hspace{-0.2in} 2 {\cal R}_{6||} + & 4 (\nabla\del \phi)_{6||} -  \frac{p-3}{2} \left({\cal R}_4  + 2 (\nabla\del \phi)_4 \right) = |H|^2 - |H|_{\bot}|^2 + e^{2\phi} \left(  |F_{k-2}|^2 - |F_{k-2}|_{\bot}|^2 \right) \label{tracepargen}\\
& + e^{2\phi} \bigg(  |F_{k}|^2 - |F_{k}|_{\bot}|^2 + |F_{k+2}|^2 + (9-p) |F_{k+4}|^2 + 5 |F_{k+6}|^2 + \frac{1}{2} (|(*_6 F_{5})|_{\bot}|^2 - |F_{5}|_{\bot}|^2) \bigg) \nn
\eea
where the $F_{5}$ terms should only be considered in IIB, we took the same conventions as for \eqref{R4T10HF}, and used that $F_{5}|_{\bot}=0$ for $p=5,7$, $F_{4}|_{\bot}=0$ for $p=6,8$, $F_{3}|_{\bot}=0$ for $p=7$, $F_{2}|_{\bot}=0$ for $p=8$. We now combine \eqref{tracepargen} with \eqref{firstsquaregen} to get
\bea
2 {\cal R}_4 + 4 (\nabla\del \phi)_4  = - &\bigg( -4 |\del \phi|^2 + 2 \Delta \phi  + \left|*_{\bot}H|_{\bot} + \varepsilon_p e^{\phi} F_{k-2}|_{\bot} \right|^2  \label{combine1}\\
& + 2 {\cal R}_{6||} + 4 (\nabla\del \phi)_{6||}  -  2 \varepsilon_p e^{\phi} (\d F_k)_{\bot} + e^{2\phi} \left( |F_{k}|^2 + |F_{k}|_{\bot}|^2 \right) \nn \\
& + e^{2\phi} \Big(  2 |F_{k+2}|^2 + (p-5) |F_{k+4}|^2 + \frac{1}{2} (|F_{5}|_{\bot}|^2 - |(*_6 F_{5})|_{\bot}|^2) \Big) \bigg)  \ . \nn
\eea
One can verify that the last line of \eqref{combine1} is always positive; we are now interested in the second line. First, we determine in Appendix \ref{ap:compute} the expression for ${\cal R}_{6||}$, which combined to the other warp factor and dilaton contributions gives
\bea
& 2 {\cal R}_4 + 4 (\nabla\del \phi)_4 - 4 |\del \phi|^2 + 2 \Delta \phi  + 4 (\nabla\del \phi)_{6||} + 2 {\cal R}_{6||} \label{use1} \\
& = 2 e^{-2A} \tilde{{\cal R}}_4 + 2 {\cal R}_{||} + 2 {\cal R}_{||}^{\bot} + \sum_{a_{||}} |(\d e^{a_{||}})|_{\bot}|^2  +2 e^{6A} \tilde{\Delta}_{\bot} e^{-4A} - 2 e^{10A} |\widetilde{\d e^{-4A}}|^2 \ .\nn
\eea
This is derived in \eqref{usefuleq} and \eqref{R6pardA0}, and the curvature terms $ {\cal R}_{||}$ and ${\cal R}_{||}^{\bot}$ are defined in \eqref{Rpar} and \eqref{Rbotpar}. We used, for simplicity, Point \ref{it:pt6} on sources in Section \ref{sec:compactif} requiring the transverse (unwarped) subspace to be a compact manifold, without boundary. Combining \eqref{use1} with $(\d F_k^{(1)})_{\bot}$ will simplify, given \eqref{squareF1} and as initially motivated there. The remaining $(\d F_k^{(0)})_{\bot}$ will combine interestingly with the warp factor terms, and $|F_{k}|_{\bot}|^2= |F_{k}^{(0)}|^2$. To that end we introduce $(G)_{\widetilde{\bot}}$, an analogous coefficient to $(G)_{\bot}$ (defined above \eqref{BI3}) on the smeared transverse subspace (i.e.~for $A=0$)
\beq
G|_{\bot}= (G)_{\widetilde{\bot}} \widetilde{{\rm vol}}_{\bot} \ ,\ (G)_{\widetilde{\bot}} = \tilde{*}_{\bot} G|_{\bot} \ ,\ (G)_{\widetilde{\bot}} = (G)_{\bot} e^{A(p-9)} \ . \label{coefsmeared}
\eeq
We then have $\tilde{\Delta}_{\bot} e^{-4A} = \tilde{*}_{\bot} \d (\tilde{*}_{\bot} \d e^{-4A}) = \left( \d (\tilde{*}_{\bot} \d e^{-4A}) \right)_{\widetilde{\bot}}$, and $ e^{\phi}  (\d F_k^{(0)})_{\bot} = e^{6A} g_s  (\d F_k^{(0)})_{\widetilde{\bot}}$. We also make use of the rewriting \eqref{trickap} that gives
\bea
e^{2\phi}|F_{k}^{(0)}|^2 = &\ e^{2\phi}| g_s^{-1} \tilde{*}_{\bot} \d e^{-4A} - \varepsilon_p  F_{k}^{(0)} |^2 + e^{10A} |\widetilde{\d e^{-4A} }|^2  \label{trick}\\
& + e^{-2A} \left(\d \left(e^{8A} \tilde{*}_{\bot} \d e^{-4A} - e^{8A} \varepsilon_p g_s F_{k}^{(0)} \right)\right)_{\widetilde{\bot}} - e^{6A} \left(\d \left( \tilde{*}_{\bot} \d e^{-4A} - \varepsilon_p g_s F_{k}^{(0)}  \right)\right)_{\widetilde{\bot}} \ . \nn
\eea
Combining \eqref{combine1} with \eqref{dFgen}, \eqref{squareF1}, \eqref{use1} and \eqref{trick} finally leads to
\begin{empheq}[innerbox=\fbox, left=\!\!\!\!\!\!\!\!\!\!\!]{align}
& 2 e^{-2A} \tilde{{\cal R}}_4  = - \left|*_{\bot}H|_{\bot} + \varepsilon_p e^{\phi} F_{k-2}|_{\bot} \right|^2  - 2 e^{2\phi}\left| g_s^{-1} \tilde{*}_{\bot} \d e^{-4A} - \varepsilon_p  F_{k}^{(0)} \right|^2 \label{FINAL} \\
& \phantom{2 e^{-2A} \tilde{{\cal R}}_4  =} - \sum_{a_{||}} \left| *_{\bot}( \d e^{a_{||}})|_{\bot} - \varepsilon_p e^{\phi} (\iota_{\del_{a_{||}}} F_k^{(1)} ) \right|^2\ - 2 {\cal R}_{||} - 2 {\cal R}_{||}^{\bot}  \nn\\
& \phantom{2 e^{-2A} \tilde{{\cal R}}_4  =} - 2 e^{-2A} \left(\d \left(e^{8A} \tilde{*}_{\bot} \d e^{-4A} - e^{8A} \varepsilon_p g_s F_{k}^{(0)} \right)\right)_{\widetilde{\bot}} \nn\\
& \ - e^{2\phi} \Big( |F_{k}|^2 - |F_{k}^{(0)}|^2 - |F_{k}^{(1)}|^2 + 2 |F_{k+2}|^2 + (p-5) |F_{k+4}|^2  + \frac{1}{2} (|F_{5}|_{\bot}|^2 - |(*_6 F_{5})|_{\bot}|^2) \Big) \nn
\end{empheq}
For clarity, we detail the last line of \eqref{FINAL}, i.e.~the $-(\mbox{flux})^2$ contribution
\bea
& p=3:\ - e^{2\phi} ( \mbox{fluxes} ) = 0 \\
& p=4:\ - e^{2\phi} ( \mbox{fluxes} ) = - 2 e^{2\phi} |F_{6}|^2 \\
& p=5:\ - e^{2\phi} ( \mbox{fluxes} ) = - e^{2\phi} \Big( |F_{3}|^2 - |F_{3}^{(0)}|^2 - |F_{3}^{(1)}|^2 + 2 |F_{5}|^2 - \frac{1}{2} |(*_6 F_{5})|_{\bot}|^2 \Big) \\
& p=6:\ - e^{2\phi} ( \mbox{fluxes} ) = - e^{2\phi} \Big( |F_{2}|^2 - |F_{2}^{(0)}|^2 - |F_{2}^{(1)}|^2 + 2 |F_{4}|^2 + |F_{6}|^2 \Big) \\
& p=7:\ - e^{2\phi} ( \mbox{fluxes} ) = - 2 e^{2\phi} \Big( |F_{3}|^2 + |F_{5}|^2 - \frac{1}{4} |(*_6 F_{5})|_{\bot}|^2 \Big)\\
& p=8:\ - e^{2\phi} ( \mbox{fluxes} ) = - e^{2\phi} \Big( 2 |F_{2}|^2 + 3 |F_{4}|^2 \Big) \ .
\eea
One has $|F_{k}|^2 - |F_{k}^{(0)}|^2 - |F_{k}^{(1)}|^2 \geq 0$ and $|F_{5}|^2 = |*_6 F_{5}|^2 \geq |(*_6 F_{5})|_{\bot}|^2$, so this line always gives a negative (semi-)definite contribution to $\tilde{\mathcal{R}}_4$.

We now integrate \eqref{FINAL} (times $\tfrac{e^{2A}}{2}$) over the six-dimensional underlying, unwarped or smeared, manifold $\widetilde{\mmm}$, considered compact (without boundary). Concretely, this amounts to multiply the expression by the unwarped six-dimensional volume form $\widetilde{{\rm vol}}_6$ and integrate. Thanks to \eqref{volrel}, this form is equal to $\widetilde{{\rm vol}}_{||} \w \widetilde{{\rm vol}}_{\bot}$. Let us focus on the total derivative term: for convenience, we denote the form under the derivative $I_{8-p} = e^{8A} \tilde{*}_{\bot} \d e^{-4A} - e^{8A} \varepsilon_p g_s F_{k}^{(0)}$. This $(8-p)$-form is along the transverse subspace. One has
\bea
& \int_{\widetilde{\mmm}} \widetilde{{\rm vol}}_6 \left(\d I_{8-p} \right)_{\widetilde{\bot}} = \int_{\widetilde{\mmm}} \widetilde{{\rm vol}}_{||} \w \left(\d I_{8-p} \right)|_{\bot}  = \int_{\widetilde{\mmm}} \widetilde{{\rm vol}}_{||} \w \d I_{8-p} = (-1)^{p} \int_{\widetilde{\mmm}} \d \widetilde{{\rm vol}}_{||} \w  I_{8-p} \nn\\
& \hspace{2.5in} = (-1)^{p+1}  \int_{\widetilde{\mmm}} \tilde{f}^{a_{||}}{}_{ b_{\bot} a_{||}}\, \tilde{e}^{b_{\bot}} \w \widetilde{{\rm vol}}_{||} \w  I_{8-p} \ .
\eea
Thanks again to Point \ref{it:pt6} on the sources in Section \ref{sec:compactif}, i.e.~the transverse (unwarped) subspace is a compact manifold without boundary, one gets $\tilde{f}^{a_{\bot}}{}_{d_{\bot}a_{\bot}} = - \tilde{f}^{a_{||}}{}_{d_{\bot}a_{||}} = 0$ (see also Appendix \ref{ap:compute}), so the above vanishes. In other words, the total derivative in \eqref{FINAL} is integrated to give zero,\footnote{Given Point \ref{it:pt6} on the sources in Section \ref{sec:compactif}, and that $\widetilde{{\rm vol}}_{\bot}$ is globally defined, one may also integrate directly over the unwarped transverse subspace. Physically, it is reasonable to assume that $I_{8-p}$ is globally defined, since it is not made of e.g.~a gauge potential but rather a flux and derivative of the warp factor; the total derivative would then be integrated to give zero. However, one may worry about the dependence of the fields on internal non-transverse directions, also in the other terms of \eqref{FINAL}. So we rather integrate over $\widetilde{\mmm}$.} resulting in
\begin{empheq}[innerbox=\fbox, left=\!\!\!\!\!\!\!\!\!\!\!\!\!\!\!]{align}
& \, \tilde{{\cal R}}_4 \int_{\widetilde{\mmm}} \widetilde{{\rm vol}}_{6}  = - \int_{\widetilde{\mmm}} \widetilde{{\rm vol}}_{6} \frac{e^{2A}}{2}  \Bigg( \left|*_{\bot}H|_{\bot} + \varepsilon_p e^{\phi} F_{k-2}|_{\bot} \right|^2  + 2 e^{2\phi}\left| g_s^{-1} \tilde{*}_{\bot} \d e^{-4A} - \varepsilon_p  F_{k}^{(0)} \right|^2 \label{FINALint} \\
& \phantom{2 \tilde{{\cal R}}_4 \int_{\widetilde{\mmm}} \widetilde{{\rm vol}}_{6}  = - \int_{\widetilde{\mmm}} \widetilde{{\rm vol}}_{6} e^{2A}} + \sum_{a_{||}} \left| *_{\bot}( \d e^{a_{||}})|_{\bot} - \varepsilon_p e^{\phi} (\iota_{\del_{a_{||}}} F_k^{(1)} ) \right|^2\ + 2 {\cal R}_{||} + 2 {\cal R}_{||}^{\bot}  \nn\\
& \ \ + e^{2\phi} \Big( |F_{k}|^2 - |F_{k}^{(0)}|^2 - |F_{k}^{(1)}|^2 + 2 |F_{k+2}|^2 + (p-5) |F_{k+4}|^2  + \frac{1}{2} (|F_{5}|_{\bot}|^2 - |(*_6 F_{5})|_{\bot}|^2) \Big)\! \Bigg) \nn
\end{empheq}

\subsection{No-go theorems}\label{sec:nogo3456}

From \eqref{FINALint}, we conclude straightforwardly on the no-go theorem
\begin{empheq}[innerbox=\fbox]{align}
& \mbox{There is no de Sitter vacuum for $p=4, 5$, or $6$, if the curvature terms vanish}  \label{nongo456}\\
& \mbox{or are positive, i.e.~for}\ {\cal R}_{||} + {\cal R}_{||}^{\bot} \geq 0 .\nn
\end{empheq}
We recall that ${\cal R}_{||}$ and ${\cal R}_{||}^{\bot}$ are defined in \eqref{Rpar} and \eqref{Rbotpar}. This no-go theorem \eqref{nongo456} is actually valid for all $3\leq p \leq 8$, as is \eqref{FINALint}. But we proved the complete absence of de Sitter vacuum for $p=7,8$ in \eqref{nogo78}, while for $p=3$, since all directions are transverse, one has by definition ${\cal R}_{||} = {\cal R}_{||}^{\bot} = 0$. This leads us to
\beq
\boxed{\mbox{There is no de Sitter vacuum for}\ p=3.} \label{nogo3}
\eeq
This result was already obtained in \cite{Blaback:2010sj}. In type IIB, the $\tilde{{\cal R}}_4$ expression \eqref{FINAL} has been obtained combining \eqref{firstsquaregen} with \eqref{traceparIIB}, with various rewritings. As indicated below \eqref{traceparIIB}, that equation is however completely vanishing for $p=3$, so one can verify that \eqref{FINAL} and \eqref{firstsquaregen} are then identical, and boil down to
\bea
p=3:\ \ 2 e^{-2A} \tilde{{\cal R}}_4   = & - \left|*_6 H + \varepsilon_3 e^{\phi} F_{3} \right|^2 - 2 \left| \tilde{*}_{6} \d e^{-4A} - \varepsilon_3 g_s  F_{5} \right|^2  \label{finalO3}\\
& -  2 e^{-2A} \left(\d \left(e^{8A} \tilde{*}_6 \d e^{-4A} - e^{8A} \varepsilon_3 g_s F_{5} \right)\right)_{\widetilde{\bot}}  \ . \nn
\eea
Integrating the above makes \eqref{nogo3} even more apparent.

Before giving more no-go theorems, let us pause and comment on ${\cal R}_{||}$ and ${\cal R}_{||}^{\bot}$ for $p=4,5,6$. These two quantities rather tend to be negative, so the no-go theorem \eqref{nongo456} would apply for them vanishing. As an example, in all known supersymmetric Minkowski vacua on twisted tori (see \cite{Andriot:2015sia}), they do vanish. ${\cal R}_{||}$ encodes the curvature of the wrapped subspace, which vanishes in the case of a flat torus. Also for $p=4$, where there is only one internal parallel direction, ${\cal R}_{||}=0$. ${\cal R}_{||}^{\bot}$ is in part encoding through $f^{a_{\bot}}{}_{b_{||}c_{||}}$ the fibration of the transverse subspace over the parallel base subspace, which is an unusual configuration. If the sources rather wrap a fiber, and the transverse subspace is a base, one can consider $\del_{a_{||}} e^{b_{\bot}}{}_m =0$, implying $f^{a_{\bot}}{}_{b_{||}c_{||}}=0$, making part of ${\cal R}_{||}^{\bot}$ vanishing. There are more instances where $f^{a_{\bot}}{}_{b_{||}c_{||}}=0$: for $p=4$, that has only one parallel direction, also for $\d \widetilde{\mbox{vol}}_{\bot}=0$, or on group manifolds where the $\tilde{f}^{a}{}_{bc}$ are constant; for the latter, the orientifold projection sets $\tilde{f}^{a_{\bot}}{}_{b_{||}c_{||}}=0$.\footnote{Generally in this paper, the orientifold projection is not helping since most objects are a priori functions and not constant, and are thus only constrained to be even or odd.} This vanishing can also be viewed as the ``T-dual'' condition to $H_{a_{||}b_{||}c_{||}}=0$, required to avoid the Freed-Witten anomaly (see e.g.~\cite{Villadoro:2007tb} and references therein), and may then be imposed.

We now turn to another important constraint on these curvatures terms for de Sitter vacua. \eqref{tracepargen} imposes
\beq
2 {\cal R}_{6||} +  4 (\nabla\del \phi)_{6||} -  \frac{p-3}{2} \left({\cal R}_4  + 2 (\nabla\del \phi)_4 \right) \geq 0 \ .\label{cond0}
\eeq
From results of Appendix \ref{ap:compute}, this quantity is found equal to $ 2 {\cal R}_{6||}|_{(\del A=0)} -  \frac{p-3}{2} e^{-2A} \tilde{{\cal R}}_4$. We deduce the following requirement for a de Sitter vacuum when $p>3$
\beq
2 {\cal R}_{6||}|_{(\del A=0)} = 2 {\cal R}_{||} + 2 {\cal R}_{||}^{\bot} + \frac{1}{2} \delta^{ch}\delta^{dj}\delta_{ab} f^{a_{||}}{}_{c_{\bot}j_{\bot}} f^{b_{||}}{}_{h_{\bot}d_{\bot}} > 0 \ ,\label{require}
\eeq
where we recall that $\sum_{a_{||}} |(\d e^{a_{||}})|_{\bot}|^2  = \frac{1}{2} \delta^{ch}\delta^{dj}\delta_{ab} f^{a_{||}}{}_{c_{\bot}j_{\bot}} f^{b_{||}}{}_{h_{\bot}d_{\bot}}$. In other words, we deduce the no-go theorem
\beq
\boxed{\mbox{There is no de Sitter vacuum for $p=4, 5$, or $6$, if}\ \, {\cal R}_{||} + {\cal R}_{||}^{\bot} < -\frac{1}{2} \sum_{a_{||}} |(\d e^{a_{||}})|_{\bot}|^2.}  \label{nogo2456}
\eeq
Note that this holds point wise. Upon integration, we can combine the requirement \eqref{require} with the one read from \eqref{FINALint}, to conclude the following
\begin{empheq}[innerbox=\fbox]{align}
& \mbox{There is no de Sitter vacuum for $p=4, 5$, or $6$, if the inequalities}  \label{constraintcurv}\\
& -\frac{1}{2} \int_{\widetilde{\mmm}} \widetilde{{\rm vol}}_{6}\, e^{2A}\,  \sum_{a_{||}} |(\d e^{a_{||}})|_{\bot}|^2\, < \, \int_{\widetilde{\mmm}} \widetilde{{\rm vol}}_{6}\, e^{2A} \left( {\cal R}_{||} + {\cal R}_{||}^{\bot} \right)\, < 0 \ \ \ \mbox{are {\it not} satisfied.}   \nn
\end{empheq}
This narrow window which would allow de Sitter can easily be checked on concrete examples.\footnote{\label{foot:p=8}For $p=8$ where there is only one transverse direction, the left-hand side of \eqref{constraintcurv} vanishes, making us recover the no-go theorem \eqref{nogo78} in that case.} The requirement $f^{a_{||}}{}_{b_{\bot}c_{\bot}}\neq0$ is particularly interesting.

We end this section with a remark on vacua T-dual to a vacuum with $O_3$. In four dimensions, the scalar potential of a gauged supergravity is invariant under T-duality (its terms and scalar fields are covariant, making the whole invariant), so its vacuum value, related to $\tilde{{\cal R}}_4$, is not changed by T-duality. In other words, from this perspective, a de Sitter vacuum does not appear by T-dualizing. Therefore, given the no-go theorem \eqref{nogo3} against de Sitter vacua with $O_3$, no T-dual vacuum to one with $O_3$ would be de Sitter either. Another argument in favor of this result goes as follows. In the case of an $O_3$, all internal directions are transverse: the (geometric) NSNS fluxes have components $H_{a_{\bot}b_{\bot}c_{\bot}}$ and $f^{a_{\bot}}{}_{b_{\bot}c_{\bot}}$. The schematic four-dimensional T-duality rule is to raise or lower indices in T-dualized directions \cite{Shelton:2005cf}, while parallel and transverse directions to a source would get exchanged. This way it is easy to generate for instance $f^{a_{||}}{}_{b_{\bot}c_{\bot}}$ from the $H$-flux. It is however impossible to generate $f^{a_{||}}{}_{b_{||}c_{||}}$, $f^{a_{\bot}}{}_{b_{\bot}c_{||}}$ or $f^{a_{\bot}}{}_{b_{||}c_{||}}$, thus leaving ${\cal R}_{||}={\cal R}_{||}^{\bot}=0$. Given \eqref{nongo456}, a vacuum T-dual to one with $O_3$ is thus not de Sitter.

\section{Outlook}\label{sec:outlook}

In this paper, we study classical de Sitter vacua of ten-dimensional type II supergravities, where the sources $D_p$ and $O_p$ have only one size $p$ and are parallel. As summarized in the Introduction, we show that there is no such de Sitter vacuum for $p=3,7,8$; for $p=4,5,6$, we cannot completely exclude these vacua, but still set high constraints on them, which amounts to having restricted values for some curvature terms of internal subspaces. These results provide clearer and tighter boundaries for the de Sitter string landscape. In addition, they can be applied concretely on various cosmological scenarios to test if those can be uplifted to string theory through a compactification. For instance, de Sitter vacua required to embed the monodromy inflation mechanism of \cite{Silverstein:2008sg} are fully excluded, completing the no-go theorem \cite{Andriot:2015aza}: indeed, this model needs an $O_4$ and $f^{a_{||}}{}_{b_{\bot}c_{\bot}}=0$, which violates the requirement \eqref{constraintcurv}. Finally, all technical tools are presented here in a self-contained manner, and can be used to pursue the search for de Sitter vacua in more involved settings.

While restrictions on the curvature terms for $p=4,5,6$ are discussed in details in Section \ref{sec:nogo3456}, one may wonder if additional information could be brought to further constrain them, and exclude completely de Sitter vacua. An idea would be to use calibration of $D_p$ and $O_p$. As discussed in Appendix \ref{ap:calib}, the conditions for calibrated sources correspond to a minimization of their energy and are thus physically relevant. Using a condition derived for sources along Minkowski \eqref{calibint}, we already obtain a no-go theorem \eqref{nogocalib}. The corresponding condition for anti-de Sitter was derived in \cite{Koerber:2007jb} and differs by a boundary term. In both cases, although not mandatory, supersymmetry serves as an interesting guideline, making the study of the de Sitter case more difficult. It would still be interesting to derive analogous conditions for de Sitter. Related geometric conditions could constrain the curvature terms further. Another idea would be to study the stability of a vacuum with such terms present. The work \cite{Danielsson:2012et} could be useful to that end: the four-dimensional scalar fields introduced there are relevant to reproduce our results in the smeared limit, and determine the stability. Proving that the curvature terms generically lead to tachyons would be an important result.

The complete exclusion of classical de Sitter vacua with parallel sources would have two important consequences. On the one hand, having parallel sources is the only setting where a complete type II supergravity description of the vacuum is possible. Indeed, having either intersecting sources, or trying to add $\NS$-sources, forces one to a partial or total smearing of the sources, at least in the current state of the art. Neglecting the backreaction of the sources in such a manner cannot always be properly justified. Progress on this is then required for any string cosmology. On the other hand, we have only focused on the shape of our universe without considering its content: matter should arise from the open string sector. In this context, the standard model would arise from intersecting branes rather than parallel branes. In addition, intersecting branes would break more supersymmetries. There is thus an optimistic view on an exclusion of classical de Sitter vacua with parallel sources: if string theory requires (specific?) intersecting branes settings to admit metastable de Sitter backgrounds, it could turn-out to be predictive when describing our universe. An application of a classical de Sitter vacuum supporting an intersecting brane model would be the description of the reheating phase after inflation. To that end, further development of intersecting brane models beyond simple torus geometries, as e.g.~in \cite{Berasaluce-Gonzalez:2016kqb}, is crucial for a connection to string cosmology.

\vspace{0.4in}

\subsection*{Acknowledgements}

We would like to thank T.~Van Riet, without whom this paper would not exist. We would also like to thank U.~H.~Danielsson, G.~Dibitetto, F.~F.~Gautason, L.~Martucci and F.~Wolf for helpful discussions before the first version, and G.~Bossard, M.~Gra\~na, K.~Krasnov, R.~Minasian, B.~Pioline, G.~Policastro, H.~Samtleben and D.~Tsimpis for useful comments that improved this draft towards its revised version. The work of D.~A.~is part of the Einstein Research Project ``Gravitation and High Energy Physics'', funded by the Einstein Foundation Berlin. The work of J.~B.~is supported by John Templeton Foundation Grant 48222, and the CEA Eurotalents program.

\newpage

\begin{appendix}
\section{Type II supergravities}\label{ap:conv}

We consider (massive) type II supergravities in string frame, supplemented with the Ramond-Ramond (RR) sources $D_p$-branes and orientifold $O_p$-planes. The bosonic part of the ten-dimensional action can be decomposed as follows
\beq
S=S_{{\rm bulk}} + S_{{\rm sources}} \ \mbox{where}\ S_{{\rm bulk}}= S_{0} + S_{CS} ,\ S_{{\rm sources}}= S_{DBI} + S_{WZ} \ .
\eeq
The bulk fields are first the metric $g_{MN}$ (${}_{M, N}$ denote ten-dimensional curved indices), the dilaton $\phi$ and the Kalb-Ramond two-form $b$. In addition, the IIA $p$-form potentials are $C_1$, $C_3$ and the IIB ones are $C_0$, $C_2$ and $C_4$. The fluxes are $H=\d b$, and the Romans mass $F_0$, $F_2=\d C_1 + b F_0 $, $F_4^{10}= \d C_3 - H \w C_1 + \frac{1}{2} b\w b F_0 $ in IIA, $F_1=\d C_0$, $F_3=\d C_2 - H\w C_0$ and $F_5^{10}$ in IIB. The corresponding action in IIA is
\beq
\hspace{-0.2in}  S_0=\frac{1}{2\kappa_{10}^2} \int \d^{10} x \, \sqrt{|g_{10}|}\ \left( e^{-2\phi} ({\cal R}_{10} +4 |\del{\phi}|^2  - \frac{1}{2} |H|^2 )
 - \frac{1}{2} (|F_0|^2 +|F_2|^2+ |F_4^{10}|^2)  \right)\ ,
\eeq
with $\ 2\kappa_{10}^2=(2\pi)^7 (\alpha^\prime)^4$, $\alpha^\prime=l_s^2$, and $|g_{10}|$ the absolute value of the determinant of the metric. For a $p$-form $A_p$, we denote $|A_p|^2 = A_{p\, M_1 \dots M_p}\, g^{M_1N_1} \! \dots g^{M_p N_p} A_{p\, N_1 \dots N_p} / p!$. In IIB, one has
\beq
\hspace{-0.2in} S_0=\frac{1}{2\kappa_{10}^2} \int \d^{10} x \, \sqrt{|g_{10}|}\ \left( e^{-2\phi} ({\cal R}_{10} +4 |\del{\phi}|^2  - \frac{1}{2} |H|^2 )
 - \frac{1}{2} (|F_1|^2 +|F_3|^2+ \frac{1}{2} |F_5^{10}|^2)  \right)\ .
\eeq
This is a pseudo-action for the flux $F_5^{10}$, that has to satisfy the following constraint on-shell
\beq
F_5^{10}= -*_{10} F_5^{10} \ . \label{constraintF5}
\eeq
The Hodge star in dimension $D$ is defined as follows, with the Levi-Civita symbol $\epsilon_{0\dots D-1}=1$,
\beq
*_{D} ( \d x^{m_1} \w \dots \w \d x^{m_p}) = \frac{\sqrt{|g_D|}}{(D-p)!}\, g^{m_1n_1} \dots g^{m_p n_p} \epsilon_{n_1 \dots n_p r_{p+1} \dots r_D} \d x^{r_{p+1}} \w \dots \w \d x^{r_D} \ . \label{Hodge}
\eeq
One has $A_p\w *_{D} A_p = \d^{D} x\, \sqrt{|g_{D}|}\ |A_p|^2 $, and we recall that $*_{D}^2 A_p = s (-1)^{p(D-p)} A_p$ for a signature $s$. From the constraint \eqref{constraintF5}, one gets on-shell
\beq
F_5^{10}\w *_{10} F_5^{10} = - *_{10} F_5^{10} \w F_5^{10} = - F_5^{10}\w *_{10} F_5^{10} \Rightarrow |F_5^{10}|^2 = 0 \ .\label{squareF5}
\eeq
This would imply that $F_5^{10}$ vanishes for a positive definite metric, which is not the case here. We will not need to specify the Chern-Simons term $S_{CS}$, so we turn to the Dirac-Born-Infeld action
\beq
S_{DBI}=- c_p\,  T_p \int_{\Sigma_{p+1}} \d^{p+1}\xi \ e^{-\phi} \sqrt{|\imath^*[g_{10} - b] + \mathcal{F}|}  \ ,
\eeq
where $\Sigma_{p+1}$ is the source world-volume and $\imath^*[ \cdot]$ the pull-back to it. The tension $T_p$ is given by $T_p^2=\frac{\pi}{\kappa_{10}^2} (4\pi^2 \alpha^\prime)^{3-p}$. For a $D_p$, $c_p =1$; for an $O_p$, $c_p =-2^{p-5}$ and $\mathcal{F}=0$. Finally, the Wess-Zumino term is given by
\beq
S_{WZ}= c_p\, \mu_p \int_{\Sigma_{p+1}} \sum_q \imath^*[C_{q}] \w e^{-\imath^*[b] + \mathcal{F}} \ ,
\eeq
where the charge $\mu_p=T_p$ for BPS sources as we consider here. One also has $\d \mathcal{F} = 0$.

We now impose two restrictions on the sources and related internal geometry that allow to promote their action to a ten-dimensional one. We first consider $-\imath^*[b] + \mathcal{F} = 0$; doing so at the level of the action instead of the equations of motion (e.o.m.) can only generate a difference in the $b$-field e.o.m.. In addition, for each source, we make the geometric considerations summarized in Point \ref{it:pt4} of Section \ref{sec:compactif} (even if the split between four and six dimensions is not required here). In particular, we use \eqref{worldvolform}, and further define the $(9-p)$-form $\delta^{\bot}_{9-p}$ that allows to remove the pull-back and promote the integral to a ten-dimensional one
\beq
S_{DBI} \stackrel{{\rm (here)}}{===} - c_p\, T_p \int e^{-\phi}\, {\rm vol}_{4} \w {\rm vol}_{||} \w \delta^{\bot}_{9-p}  \ , \ \ S_{WZ} \stackrel{{\rm (here)}}{===}  c_p\, \mu_p \int C_{p+1} \w \delta^{\bot}_{9-p}  \ ,\label{WZ10D}
\eeq
where the form ordering is a convention choice. Given the volume forms relations \eqref{volrel}, $\delta^{\bot}_{9-p}$ is understood as given by ${\rm vol}_{\bot}$ divided by the transverse metric determinant, times a formal delta function $\delta(\bot)$ that localizes the source in the transverse directions. We also introduce a projector $P[ \cdot]$ to the source (parallel) directions, giving ${\rm vol}_{4} \w {\rm vol}_{||} \w \delta^{\bot}_{9-p} = \d^{10} x \sqrt{|P[g_{10}]|} \delta(\bot)$; this rewriting is more convenient.

We now derive the Einstein equation and dilaton e.o.m.. $S_{CS}$ and $S_{WZ}$ are topological terms that do not depend on $g_{MN}$ or $\phi$, so they do not contribute. We define the energy momentum tensor as
\beq
\frac{1}{\sqrt{|g_{10}|}} \sum_{{\rm sources}} \frac{\delta S_{DBI}}{\delta g^{MN}}\equiv - \frac{e^{-\phi}}{4 \kappa_{10}^2} T_{M N} \ .
\eeq
It is given here, together with its trace, by
\bea
& T_{M N}= - \frac{2\kappa_{10}^2}{\sqrt{|g_{10}|}} \sum_{{\rm sources}} c_p\,  T_p\ P[g_{MN}] \sqrt{|P[g_{10}]|} \ \delta(\bot) \ ,\\
& T_{10}= g^{MN} T_{M N} = - \frac{2\kappa_{10}^2}{\sqrt{|g_{10}|}} \sum_{{\rm sources}} c_p\,  T_p\ (p+1) \sqrt{|P[g_{10}]|} \ \delta(\bot) \equiv \sum_{{\rm sources}}  (p+1) \ t_p \ .\label{T10}
\eea
One can then verify
\beq
\frac{1}{\sqrt{|g_{10}|}} \sum_{{\rm sources}} \frac{\delta S_{DBI}}{\delta \phi}=- \frac{e^{- \phi}}{2 \kappa_{10}^2} \sum_{{\rm sources}}  t_p \ .
\eeq
We deduce the dilaton equation of motion and the Einstein equation\footnote{On the dilaton terms in the Einstein equation, we refer to Footnote 30 of \cite{Andriot:2011iw}. We also recall the Laplacian on a function $\varphi$: $\Delta \varphi = g^{MN} \nabla_M \del_N \varphi = \frac{1}{\sqrt{|g|}} \del_M (\sqrt{|g|} g^{MN} \del_N \varphi )$; $\Delta$ stands here for the ten-dimensional one.} in type IIA and IIB
\bea
2 {\cal R}_{10} -|H|^2 + 8( & \Delta \phi - |\del \phi|^2 ) = -e^{\phi} \sum_{{\rm sources}}  t_p \ ,\label{dileom}\\
{\cal R}_{MN}-\frac{g_{MN}}{2} {\cal R}_{10} & = \frac{1}{4} H_{MPQ}H_N^{\ \ PQ}+\frac{e^{2\phi}}{2}\left(F_{2\ MP}F_{2\ N}^{\ \ \ \ P} +\frac{1}{3!} F^{10}_{4\ MPQR}F_{4\ N}^{10 \ \ PQR} \right) \nn\\
& + \frac{e^{\phi}}{2}T_{MN} -\frac{g_{MN}}{4} \left( |H|^2 + e^{2\phi}(|F_0|^2 + |F_2|^2 + |F_4^{10}|^2 )\right) \label{EinstIIA}\\
& -2\nabla_M \del_N{\phi} +2 g_{MN} (\Delta \phi - |\del \phi|^2 ) \nn   \ ,  \\
{\cal R}_{MN}-\frac{g_{MN}}{2} {\cal R}_{10} & = \frac{1}{4} H_{MPQ}H_N^{\ \ PQ}+\frac{e^{2\phi}}{2}\left(F_{1\ M}F_{1\ N} +\frac{1}{2!} F_{3\ MPQ}F_{3\ N}^{\ \ \ \ PQ} +\frac{1}{2\cdot4!} F_{5\ MPQRS}^{10}F_{5\ N}^{10 \ \ PQRS} \right) \nn\\
& + \frac{e^{\phi}}{2}T_{MN} -\frac{g_{MN}}{4} \left( |H|^2 + e^{2\phi}(|F_1|^2 + |F_3|^2 )\right) \label{EinstIIB}\\
& -2\nabla_M \del_N{\phi} +2 g_{MN} (\Delta \phi - |\del \phi|^2 ) \nn   \ ,
\eea
where we imposed the constraint \eqref{squareF5}.

We now turn to the fluxes. As pointed-out in the seminal paper \cite{Bergshoeff:2001pv}, the Wess-Zumino action \eqref{WZ10D} is problematic for the higher $D_{p}$-branes, and the magnetic coupling. To derive the fluxes Bianchi identities (BI) and e.o.m.~in presence of sources, one should then use the democratic formalism, that has in addition the advantage of avoiding the Chern-Simons terms. One replaces the previous RR action for the following pseudo-action
\beq
\frac{1}{2 \kappa_{10}^2} \int \left( - \frac{1}{4}\right) \sum_{q}
F_{q+1} \w *_{10} F_{q+1} + \sum_{{\rm sources},\, q} c_{q-1} \mu_{q-1} \int \frac{1}{2} C_q \w \delta^{\bot}_{10-q} \ ,
\eeq
where $q=1, 3, 5, 7, 9$ for IIA and $q=0, 2, 4, 6, 8$ for IIB, with $F_p=\d C_{p-1} - H\w C_{p-3} + F_0 e^b|_p$ consistently with above. One should then impose on-shell the following constraint
\beq
F_p = (-1)^{\left[\frac{p+1}{2} \right]} *_{10} F_{10-p} \ , \label{constraint}
\eeq
where the integer part of $\frac{p}{2}$ can be rewritten as $(-1)^{\left[\frac{p}{2} \right]}=(-1)^{\frac{p(p-1)}{2}}$, giving $(-1)^{\left[\frac{p+1}{2} \right]}= (-1)^{\frac{(p+1)p}{2}} = (-1)^p (-1)^{\left[\frac{p}{2} \right]}$. The e.o.m.~for $C_q$ is now
\beq
\d ( *_{10} F_{q+1}) + H \w *_{10} F_{q+3} = 2 \kappa_{10}^2\ (-1)^{q+1} \!\!\! \sum_{(q-1)-{\rm sources}} \!\!\! c_{q-1} \mu_{q-1}\, \delta^{\bot}_{10-q} \ .\label{eom}
\eeq
Imposing the constraint gives the equivalent equation
\beq
\d (F_{9-q}) - H\w F_{7-q} = 2 \kappa_{10}^2\ (-1)^{q+1} \!\!\! \sum_{(q-1)-{\rm sources}} \!\!\! c_{q-1} \mu_{q-1}\, \lambda(\delta^{\bot}_{10-q})\ ,\label{BI}
\eeq
where for a $p$-form $A_p$, $\lambda(A_p)=(-1)^{\left[\frac{p}{2} \right]} A_p$. To get respectively the standard e.o.m.~and BI, we restrict to the standard fluxes, giving in IIA and IIB
\bea
& \d ( *_{10} F_2) + H \w *_{10} F_{4}^{10} = 2 \kappa_{10}^2 \!\!\! \sum_{0-{\rm sources}} \!\!\! c_{0} \mu_{0}\, \delta^{\bot}_{9} \ ,\ \ \d ( *_{10} F_{4}^{10}) + H \w F_{4}^{10} = 2 \kappa_{10}^2 \!\!\! \sum_{2-{\rm sources}} \!\!\! c_{2} \mu_{2}\, \delta^{\bot}_{7} \ ,\nn\\
& \d (F_{0}) = 2 \kappa_{10}^2 \!\!\! \sum_{8-{\rm sources}} \!\!\! c_{8} \mu_{8}\, \delta^{\bot}_{1}\ ,\ \ \d (F_{2}) - H\w F_{0} = - 2 \kappa_{10}^2 \!\!\! \sum_{6-{\rm sources}} \!\!\! c_{6} \mu_{6}\, \delta^{\bot}_{3}\ , \\
& \d (F_{4}^{10}) - H\w F_{2} = 2 \kappa_{10}^2 \!\!\! \sum_{4-{\rm sources}} \!\!\! c_{4} \mu_{4}\, \delta^{\bot}_{5}\ ;\nn\\
& \d ( *_{10} F_{1}) + H \w *_{10} F_{3} = 0 \ ,\ \ \d ( *_{10} F_{3}) + H \w *_{10} F_{5}^{10} = - 2 \kappa_{10}^2 \!\!\! \sum_{1-{\rm sources}} \!\!\! c_{1} \mu_{1}\, \delta^{\bot}_{8} \ ,\\
& \d ( *_{10} F_{5}^{10}) + H \w F_{3} = - 2 \kappa_{10}^2 \!\!\! \sum_{3-{\rm sources}} \!\!\! c_{3} \mu_{3}\, \delta^{\bot}_{6} \ \Leftrightarrow\ \d ( F_{5}^{10}) - H \w F_{3} = 2 \kappa_{10}^2 \!\!\! \sum_{3-{\rm sources}} \!\!\! c_{3} \mu_{3}\, \delta^{\bot}_{6} \ ,\nn\\
& \d (F_{1}) = 2 \kappa_{10}^2 \!\!\! \sum_{7-{\rm sources}} \!\!\! c_{7} \mu_{7}\, \delta^{\bot}_{2} \ ,\ \ \d (F_{3}) - H\w F_{1} = - 2 \kappa_{10}^2 \!\!\! \sum_{5-{\rm sources}} \!\!\! c_{5} \mu_{5}\, \delta^{\bot}_{4} \ .\nn
\eea
Finally, with the above pseudo-action, the $b$-field e.o.m.~is given by
\beq
\d (e^{-2\phi} *_{10}H) -  \sum_{1\leq q \leq 4} F_{q-1} \w *_{10} F_{q+1} - \frac{1}{2} F_4^{10} \w F_4^{10} = \mbox{sources}\ ,
\eeq
where the democratic formalism constraint \eqref{constraint} has been applied. The right-hand side ``sources'' denotes collectively the contribution from $S_{{\rm sources}}$ as well as from the source term in the right-hand side of \eqref{eom} that has been used. The latter seems to cancel the contribution from $S_{WZ}$, leaving only the contribution from $S_{DBI}$, as pointed-out in \cite{Koerber:2007hd}. We will not use this e.o.m.. In absence of $\NS_5$-branes as here, the BI is
\beq
\d H=0\ .
\eeq

\subsection*{Relation to other conventions}

We follow the democratic formalism conventions \cite{Bergshoeff:2001pv} (the same as \cite{Grana:2006kf}), except for the Hodge star definition, where we get a sign $(-1)^{(D-p)p}=(-1)^{(D-1)p}$. We thus make this sign explicit, as in the constraint \eqref{constraint}. Also, in \cite{Bergshoeff:2001pv} is not considered a $b$-field in the sources action, for which we then follow consistently \cite{Bergshoeff:1996tu}.

Another set of conventions in the literature are those of e.g.~\cite{Martucci:2005ht, Koerber:2007hd, Lust:2008zd, Koerber:2010bx}. These conventions differ from the democratic formalism ones by a change of sign of $H$ in IIB, and the change $C_q \rightarrow (-1)^{\frac{q-1}{2}} C_q$ in IIA. The latter is equivalent to {\it no change of $C_q$}, but a change of sign of $H$ together with $F_{q+1} \rightarrow (-1)^{\frac{q-1}{2}} F_{q+1}$ for $q\geq0$, leading to rewriting the constraint by replacing $(-1)^{\left[\frac{p}{2} \right]}$ by $(-1)^{\left[\frac{p-1}{2} \right]}$. That replacement is neutral in IIB, so the constraint can be rewritten in both theories. The map from our conventions to those of \cite{Martucci:2005ht, Koerber:2007hd, Lust:2008zd, Koerber:2010bx}, for both IIA and IIB, is then to change the sign of $H$, or actually of the $b$-field, rewriting the constraint \eqref{constraint} by replacing $(-1)^{\left[\frac{p}{2} \right]} \rightarrow (-1)^{\left[\frac{p-1}{2} \right]}$, and changing the Hodge star by an appropriate sign.\footnote{Another difference is the value of the Levi-Civita symbol, which is opposite. This has no impact here since this symbol is considered only formally, defining e.g.~$\d^{10} x \equiv 1/10! \ \epsilon_{M_1 \dots M_{10}} \d x^{M_1} \w \dots \w \d x^{M_{10}}$. It may matter if one computes explicit duals of forms, as e.g.~the RR fluxes in \cite{Grana:2006kf, Andriot:2015sia}.} Upon this map, one can verify that the e.o.m.~\eqref{eom} and (B.6) of \cite{Koerber:2007hd} match, using $\delta^{\bot}_{10-(p+1)}= \lambda(j_{(\Sigma_p,0)})$. But since the constraint differs by a sign, the BI differ by a sign (in IIA), making them not equivalent. That sign has however no physical relevance: it could be avoided by changing the sign of the WZ term in the brane action, which amounts to change the definition of brane versus anti-brane, or equivalently change which of the two type II supersymmetries are preserved (and its projector), or change the orientation of the world-volume. Note that the calibration (poly)form, related to the volume form of the sources, would then also pick a sign in IIA.

\section{$D_p$-brane energy minimization and calibration}\label{ap:calib}

We derive in this appendix conditions to minimize the energy of a $D_p$-brane, for a Minkowski four-dimensional space-time. They should correspond to e.o.m. of the $D_p$-brane own bosonic degrees of freedom, namely of its scalar fields,\footnote{The other $D_p$-brane degrees of freedom are the gauge potentials and flux $\mathcal{F}$, but having them to vanish is a solution to their e.o.m., at least for Minkowski, so we do not need to consider this other e.o.m. further.} and should thus be satisfied when looking for a vacuum. Minimizing the $D_p$-brane energy is related to the notion of calibration, that we will first recall. Most of the work on this topic has been made for supersymmetric vacua; we go here beyond this context as we do not consider supersymmetry.

We first follow \cite{Martucci:2005ht} and consider a $D_p$-brane on a world-volume $\Sigma_{p+1}$ with flux ${\cal F}$. A generalized calibration \cite{Koerber:2005qi, Martucci:2005ht, Koerber:2006hh, Koerber:2007hd, Witt, Martucci:2011dn}, denoted $\omega$, is a sum of forms of different degrees, or polyform, in ten dimensions such that $(\d - H\w)\omega = 0$ and
\beq
\imath^*[\omega] \w e^{-\imath^*[b] + \mathcal{F}} \leq {\cal E}(\Sigma_{p+1}, \mathcal{F}) \ ,\label{calibration}
\eeq
where the inequality is understood under projection on $\Sigma_{p+1}$ along $\d^{p+1}\xi$. The coefficient of ${\cal E}$ gives the energy density. For static configurations (as considered in this paper), ${\cal E}$ is read from $S_{DBI} + S_{WZ}$ and is given, for $T_p=\mu_p$, by\footnote{The time direction may need to be removed from the forms; we refer to \cite{Martucci:2005ht} or Appendix A of \cite{Koerber:2007jb} for more details.}
\beq
{\cal E}(\Sigma_{p+1}, \mathcal{F}) = c_p\,  T_p \left(  \d^{p+1}\xi \ e^{-\phi} \sqrt{|\imath^*[g_{10} - b] + \mathcal{F}|} - \sum_q \imath^*[C_{q}] \w e^{-\imath^*[b] + \mathcal{F}} \right) \ .
\eeq
A $D_p$-brane is said to be calibrated by $\omega$, in a generalized sense, if $\omega$ saturates the inequality for some $\Sigma_{p+1}$ and $\mathcal{F}$
\beq
\imath^*[\omega] \w e^{-\imath^*[b] + \mathcal{F}} = {\cal E}(\Sigma_{p+1}, \mathcal{F}) \ .\label{calibrated}
\eeq
As argued in \cite{Martucci:2005ht}, under some conditions, this saturation can be understood as a minimization of the energy $E$. Indeed, for $\Sigma_{p+1}$ being a cycle, consider continuous deformations to $\Sigma_{p+1}'$ in the same homology class, i.e.~$\Sigma_{p+1}' - \Sigma_{p+1} = \del {\cal B}$, with $\hat{\mathcal{F}}$ on ${\cal B}$ restricting to $\mathcal{F}$ or $\mathcal{F}'$. Then
\bea
E(\Sigma_{p+1}',\mathcal{F}') & = \int_{\Sigma_{p+1}'} {\cal E}(\Sigma_{p+1}', \mathcal{F}') \label{reasoning}\\
& \geq \int_{\Sigma_{p+1}'} \imath^*[\omega] \w e^{-\imath^*[b] + \mathcal{F}'} = \int_{\Sigma_{p+1}} \imath^*[\omega] \w e^{-\imath^*[b] + \mathcal{F}} + \int_{{\cal B}} \imath^*[(\d - H\w)\omega] \w e^{-\imath^*[b] + \hat{\mathcal{F}}}\nn\\
&  = \int_{\Sigma_{p+1}} \imath^*[\omega] \w e^{-\imath^*[b] + \mathcal{F}} = \int_{\Sigma_{p+1}} {\cal E}(\Sigma_{p+1}, \mathcal{F})\nn\\
& = E(\Sigma_{p+1},\mathcal{F}) \ .\nn
\eea
So a calibrated $D_p$-brane has its energy minimized. Note that the fluctuations of $\Sigma_{p+1}$ can be understood as that of the embedding coordinates, i.e.~the brane scalar fields. The above reasoning holds e.g.~for $\Sigma_{p+1}$ being the product of Minkowski times an internal cycle \cite{Martucci:2005ht}.

Given the equality \eqref{calibrated}, a candidate for the calibration $\omega$ is given by
\beq
\omega= c_p T_p \left( e^{-\phi}\, {\rm vol}_{\Sigma_{p+1}, \mathcal{F}} \w e^{b - \mathcal{F}} - \sum_{q\leq p+1} C_{q}|_4 \right) \ ,
\eeq
where the form ${\rm vol}_{\Sigma_{p+1}, \mathcal{F}}$ is such that $\imath^*[{\rm vol}_{\Sigma_{p+1}, \mathcal{F}}] = \d^{p+1}\xi \ \sqrt{|\imath^*[g_{10} - b] + \mathcal{F}|}$, and we restrict to the components of $C_q$ containing the full four-dimensional volume form. For space-time filling sources, $\imath^*[C_{q}|_4]= \imath^*[C_{q}]$ so this restriction is only future convenience; the same holds for $q\leq p+1$. A minimized energy becomes equivalent to $(\d - H\w)\omega = 0$, i.e.~to
\bea
& F_{h < p+2}\,|_{4} = 0 \ , \label{dE=01}\\
& \d\left( e^{-\phi}\, {\rm vol}_{\Sigma_{p+1}, \mathcal{F}} \right) - F_{p+2}|_{4} =0 \ , \label{dE=02}
\eea
with the RR fluxes restricted to their components containing the four-dimensional volume.\footnote{One would naively get on top of \eqref{dE=01}, \eqref{dE=02}, equations on forms of higher degree. But the projection in \eqref{calibrated} and \eqref{reasoning} actually bounds the relevant degrees of $\omega$ and $(\d - H\w)\omega$, avoiding higher degrees with a refined reasoning.}

We now consider $D_p$-branes having the properties of Points \ref{it:pt1}, \ref{it:pt2}, \ref{it:pt4} of Section \ref{sec:compactif}. In particular, $-\imath^*[b] + \mathcal{F} = 0$ gives with \eqref{worldvolform} ${\rm vol}_{\Sigma_{p+1}, \mathcal{F}} = {\rm vol}_{4} \w {\rm vol}_{||}$. The metric \eqref{10dmetric} allows to extract the unwarped four-dimensional volume form as ${\rm vol}_{4}= e^{4A} \widetilde{{\rm vol}}_4$. On top, we have used the electric RR coupling to the $D_p$-brane while for us $p\geq 3$: as explained in Appendix \ref{ap:conv}, this requires to use the democratic formalism. Because $p\geq 3$, the $F_{p+2}$ are the higher fluxes only: to recover proper fluxes of type II supergravities, we then have to use the democratic formalism constraint \eqref{constraint} (and sources become the magnetic ones). We rewrite \eqref{dE=02} as
\bea
& \widetilde{{\rm vol}}_4 \w \d\left( e^{4A -\phi} {\rm vol}_{||} \right) = (-1)^{\left[\frac{p+3}{2} \right]} (*_{10} F_{8-p})|_{4} = (-1)^{\left[\frac{p+3}{2} \right]} *_{10} (F_{8-p})|_{\subset 6} \nn\\
\Leftrightarrow \ & e^{-4A} *_6 \d\left( e^{4A -\phi} {\rm vol}_{||} \right) = (-1)^{\left[\frac{p+3}{2} \right]} (-1)^p\ F_{8-p} \ , \nn
\eea
where from the last line on, we drop the projection on internal components $|_{\subset 6}$, since $F_k$ without any index ${}^{4}$ or ${}^{10}$ denotes the internal part of the flux. Similarly, we rewrite \eqref{dE=01}
\beq
0 = (*_{10} F_{l > 8-p})|_{4} = *_{10} (F_{l > 8-p})|_{\subset 6} = *_{10} F_{l > 8-p} \ .
\eeq
Using that $(-1)^{\left[\frac{p+3}{2} \right]} (-1)^p = - (-1)^{\left[\frac{p}{2} \right]} = - (-1)^{\left[\frac{8-p}{2} \right]} (-1)^{8-p} = - (-1)^{\left[\frac{9-p}{2} \right]}=(-1)^{p} \varepsilon_p$, we rewrite the above as
\bea
& F_{k} = (-1)^{p} \varepsilon_p e^{-4A} *_6 \d\left( e^{4A -\phi} {\rm vol}_{||} \right) \ ,\quad 0 \leq k=8-p \leq 5 \ , \label{Fkcalibap}\\
& F_{l > k} = 0 \ . \label{Fl>k}
\eea
We call \eqref{Fkcalibap} the calibration condition. As shown in \cite{Martucci:2005ht}, it is automatically satisfied for a supersymmetric Minkowski background, using the supersymmetry preserved by the $D_p$-brane. Here, we considered Minkowski and conditions on the sources, but not supersymmetry. We proved that minimizing the energy was equivalent to \eqref{Fkcalibap} and \eqref{Fl>k}.

In the case of a four-dimensional anti-de Sitter space-time, \eqref{Fkcalibap} gets modified by an additional term $X$, as explained in \cite{Koerber:2007jb}. This can be seen already through the supersymmetry conditions for anti-de Sitter, that include $X$. This term is due to the space boundaries of anti-de Sitter \cite{Koerber:2007jb}; those require to adapt the reasoning \eqref{reasoning}, not valid otherwise. Here, the question would be to determine analogous conditions for de Sitter. We cannot be guided by supersymmetry, but we still expect a correction to \eqref{Fkcalibap}, related to properties of de Sitter space-time. Getting such conditions would bring relevant new constraints for de Sitter vacua.

\section{Computational details}\label{ap:compute}

In this appendix, we compute various terms involving the warp factor and the dilaton, as well as curvature terms. We first compute the ten-dimensional Ricci tensor in curved indices along four-dimensional directions, and the corresponding scalar ${\cal R}_4= g^{MN} {\cal R}_{MN=\mu\nu} $, for the Levi-Civita connection; see e.g.~\cite{Andriot:2013xca, Andriot:2015aza} for relevant formulas. We extract the warp factor dependence using the metric \eqref{10dmetric} and obtain
\bea
{\cal R}_{MN=\mu\nu} & =  \tilde{{\cal R}}_{\mu\nu}  - \frac{1}{2} \tilde{g}_{\mu\nu} \left( \Delta_6\, e^{2A} + e^{-2A} (\del e^{2A})^2 \right)  \label{Riccipar} \ ,\\
{\cal R}_4  & = e^{-2A} \tilde{{\cal R}}_4 -2 e^{-2A} \left( \Delta_6\, e^{2A} + e^{-2A} (\del e^{2A})^2 \right) \ , \label{R4}
\eea
where $\tilde{{\cal R}}_{\mu\nu}$ is the purely four-dimensional Ricci tensor built from $\tilde{g}_{\mu\nu}$ and $\tilde{{\cal R}}_4$ its Ricci scalar, $\Delta_6$ is the internal Laplacian and $(\del e^{2A})^2 = g^{mn} \del_m e^{2A} \del_n e^{2A} $. This computation required the following connection coefficient
\beq
\Gamma^{P=p}_{MN=\mu\nu} = - \frac{1}{2} \tilde{g}_{\mu\nu} g^{pn}  \del_{n} e^{2A}  \ , \label{Chris}
\eeq
that we use again to compute the quantities
\bea
2 (\nabla\del \phi)_4 \equiv 2 g^{MN=\mu\nu} \left(\del_M \del_N \phi - \Gamma^{P}_{MN} \del_P \phi  \right) = -2 e^{2\phi-2A} g^{mn}  \del_{m} e^{2A} \del_{n} e^{-2\phi} \ , \label{Ndphi4}\\
4 |\del \phi|^2 - 2 \Delta \phi=  e^{2\phi} \Delta e^{-2\phi} = e^{2\phi} \Delta_6\, e^{-2\phi} + 2 e^{2\phi-2A} g^{mn} \del_{m} e^{2A} \del_{n} e^{-2\phi} \ , \label{Deltaphi}
\eea
where the second equations made use that the dilaton depends only on internal directions.

We now use these results in equation \eqref{4dtracefinal}: it involves the quantity
\bea
& (p-3) {\cal R}_4 - 2 e^{2\phi} \Delta e^{-2\phi} + 2 (p-3) (\nabla\del \phi)_4 \label{qttyap}\\
=&\, (p-3) e^{-2A} \tilde{{\cal R}}_4 - 2 e^{2\phi -2A(p-3)} \Delta_6\, e^{2A(p-3)-2\phi} + 2(p-5) e^{2\phi -2A(p-2)} g^{mn} \del_m e^{2A} \del_n e^{2A(p-3)-2\phi} \ .\nn
\eea
The last two terms vanish for the standard value of a $D_p$-brane solution, picked in this paper
\beq
e^{\phi}= g_s e^{A(p-3)} \label{dilatonwarpap} \ ,
\eeq
where $g_s$ is a constant.\footnote{The value \eqref{dilatonwarpap} might be derived rather than imposed, as we sketch here. Using more knowledge on the internal metric, one may show that the last two terms of \eqref{qttyap} actually combine into one term with a Laplacian of the unwarped internal metric $\tilde{\Delta}_6\, e^{2A(p-3)-2\phi}$. For $p=7, 8$, this quantity can have a definite sign through \eqref{4dtracefinal}. Using an integration, one can then show that this quantity vanishes. Harmonic functions on a compact manifold without boundary are constant, so one would derive this way \eqref{dilatonwarpap}.} We conclude that the quantity \eqref{qttyap} is equal to $(p-3) e^{-2A} \tilde{{\cal R}}_4$, and use this result in Section \ref{sec:O7O8}.

We now turn to \eqref{combine1} that involves the quantity
\beq
2 {\cal R}_4 + 4 (\nabla\del \phi)_4 - 4 |\del \phi|^2 + 2 \Delta \phi  + 4 (\nabla\del \phi)_{6||} + 2 {\cal R}_{6||} \ . \label{qtty0ap}
\eeq
To compute the last two terms, we need to use flat indices (definitions in Section \ref{sec:compactif} and around \eqref{fabc}). For the Levi-Civita connection, the spin connection is related to $f^{a}{}_{bc}$, so that one has generically for the Ricci tensor in flat indices (see e.g.~\cite{Andriot:2014uda})
\bea
& 2\ {\cal R}_{cd} = \del_a f^a{}_{cd} + 2 \eta^{ab} \del_a f^g{}_{b(c} \eta_{d)g} - 2 \del_c f^b{}_{bd} \label{Ricciflat} \\
&\phantom{2\ {\cal R}_{cd} } + f^a{}_{ab} \left(f^b{}_{cd} + 2 \eta^{bg} f^h{}_{g(c} \eta_{d)h} \right) - f^b{}_{ac} f^a{}_{bd} - \eta^{bg} \eta_{ah} f^h{}_{gc} f^a{}_{bd} + \frac{1}{2} \eta^{ah}\eta^{bj}\eta_{ci}\eta_{dg} f^i{}_{aj} f^g{}_{hb} \ .\nn
\eea
From now on, we denote respectively ${}_A, {}_{\alpha}, {}_a$ the flat ten-, four- and six-dimensional indices, and refer to the metric \eqref{metricwarp}. We take $\del_{\alpha} A = 0 \ ,\  \del_{a_{||}} A = 0$, which implies
\beq
f^{a_{||}}{}_{BC}= \delta^b_B \delta^c_C f^{a_{||}}{}_{bc} \ ,\ f^A{}_{Bc_{||}}= \delta^A_a \delta^b_B f^a{}_{bc_{||}} \ , \ f^A{}_{Bc_{\bot}}= \delta^A_a \delta^b_B f^a{}_{bc_{\bot}} + \delta^A_{\alpha} \delta^{\beta}_B \delta^{\alpha}_{\beta} e^{-A} \del_{c_{\bot}} e^A \ .
\eeq
This allows to compute ${\cal R}_{6||}=\eta^{AB} {\cal R}_{AB=a_{||}b_{||}}$, giving
\bea
2 {\cal R}_{6||}  = &\ 2 \delta^{cd} \del_c f^{a_{||}}{}_{da_{||}} +8 \delta^{cd} f^{a_{||}}{}_{c_{\bot}a_{||}} e^{-A} \del_{d_{\bot}} e^A  - 2 \delta^{ab} \del_{a_{||}} f^{c}{}_{cb_{||}} + 2 \delta^{cd} f^{a_{||}}{}_{ca_{||}}  f^{e}{}_{ed} \nn\\
& -\delta^{ab} f^d{}_{ca_{||}} f^c{}_{db_{||}} - \delta^{ab} \delta^{dg} \delta_{ch} f^h{}_{ga_{||}} f^c{}_{db_{||}} + \frac{1}{2} \delta^{ch}\delta^{dj}\delta_{ab} f^{a_{||}}{}_{cj} f^{b_{||}}{}_{hd} \label{Ricciflatpartrace1}\\
=& \ 2 \delta^{cd} \del_{c_{\bot}} f^{a_{||}}{}_{d_{\bot}a_{||}} +8 \delta^{cd} f^{a_{||}}{}_{c_{\bot}a_{||}} e^{-A} \del_{d_{\bot}} e^A   - 2 \delta^{ab} \del_{a_{||}} f^{c}{}_{cb_{||}} + 2 \delta^{cd} f^{a_{||}}{}_{ca_{||}}  f^{e}{}_{ed} \nn\\
& + 2 {\cal R}_{||} + 2 {\cal R}_{||}^{\bot} + \frac{1}{2} \delta^{ch}\delta^{dj}\delta_{ab} f^{a_{||}}{}_{c_{\bot}j_{\bot}} f^{b_{||}}{}_{h_{\bot}d_{\bot}} \ ,\label{Ricciflatpartrace2}\\
\mbox{where}\ \ 2 {\cal R}_{||}  =&\ 2 \delta^{cd} \del_{c_{||}} f^{a_{||}}{}_{d_{||}a_{||}} -\delta^{ab} f^{d_{||}}{}_{c_{||}a_{||}} f^{c_{||}}{}_{d_{||}b_{||}}- \frac{1}{2} \delta^{ch}\delta^{dj}\delta_{ab} f^{a_{||}}{}_{c_{||}j_{||}} f^{b_{||}}{}_{h_{||}d_{||}} \ , \label{Rpar}\\
2 {\cal R}_{||}^{\bot}  =&  -\delta^{ab} f^{d_{\bot}}{}_{c_{\bot}a_{||}} f^{c_{\bot}}{}_{d_{\bot}b_{||}} - \delta^{ab} \delta^{dg} \delta_{ch} f^{h_{\bot}}{}_{g_{\bot}a_{||}} f^{c_{\bot}}{}_{d_{\bot}b_{||}} \label{Rbotpar}\\
& - 2 \delta^{ab} f^{d_{\bot}}{}_{c_{||}a_{||}} f^{c_{||}}{}_{d_{\bot}b_{||}} - \delta^{ab} \delta^{dg} \delta_{ch} f^{h_{\bot}}{}_{g_{||}a_{||}} f^{c_{\bot}}{}_{d_{||}b_{||}} \ .\nn
\eea
We now extract the warp factor with $e^{a_{||}}{}_m=e^A \tilde{e}^{a_{||}}{}_m$, $e^{a_{\bot}}{}_m=e^{-A} \tilde{e}^{a_{\bot}}{}_m$. We first obtain
\bea
& f^{a_{||}}{}_{b_{||}c_{||}}=e^{-A} \tilde{f}^{a_{||}}{}_{b_{||}c_{||}} \ ,\ f^{a_{\bot}}{}_{b_{||}c_{||}}=e^{-3A} \tilde{f}^{a_{\bot}}{}_{b_{||}c_{||}} \ ,\ f^{a_{\bot}}{}_{b_{\bot}c_{||}}=e^{-A} \tilde{f}^{a_{\bot}}{}_{b_{\bot}c_{||}} \ ,\\
& f^{a_{||}}{}_{b_{\bot}c_{||}}=e^{A} \tilde{f}^{a_{||}}{}_{b_{\bot}c_{||}} - \delta^{a_{||}}_{c_{||}} \del_{\tilde{b}_{\bot}} e^{A} \ ,\ f^{a_{\bot}}{}_{b_{\bot}c_{\bot}} = e^A \tilde{f}^{a_{\bot}}{}_{b_{\bot}c_{\bot}} + 2 e^{2A} \delta^{\tilde{a}_{\bot}}_{[\tilde{b}_{\bot}} \del_{\tilde{c}_{\bot}]} e^{-A} \ .\nn
\eea
With in addition $f^{a_{||}}{}_{c_{\bot}j_{\bot}}=e^{3A}\tilde{f}^{a_{||}}{}_{c_{\bot}j_{\bot}}$, one shows that the last line in \eqref{Ricciflatpartrace2} does not produce any $\del A$, so contributes to what we denote ${\cal R}_{6||}|_{(\del A=0)}$. We turn to the other line. For a compact manifold (without boundary), one generically has $f^a{}_{ab}=0$. Here the relevant manifold is the unwarped or smeared one (see above \eqref{metricwarp}), meaning the correct condition is $\tilde{f}^a{}_{ab}=0$. From the above, we deduce
\beq
f^a{}_{ab_{||}} = 0 \ ,\ f^a{}_{ab_{\bot}} = (2p-11) \del_{\tilde{b}_{\bot}} e^A \ .
\eeq
We then compute the first line of \eqref{Ricciflatpartrace2}
\bea
& 2 \delta^{cd} \del_{c_{\bot}} f^{a_{||}}{}_{d_{\bot}a_{||}} +8 \delta^{cd} f^{a_{||}}{}_{c_{\bot}a_{||}} e^{-A} \del_{d_{\bot}} e^A   - 2 \delta^{ab} \del_{a_{||}} f^{c}{}_{cb_{||}} + 2 \delta^{cd} f^{a_{||}}{}_{ca_{||}}  f^{e}{}_{ed}\\
=\ & 2 \delta^{cd} e^{2A} \del_{\tilde{c}_{\bot}} \tilde{f}^{a_{||}}{}_{d_{\bot}a_{||}} +  e^{2A(p-2)} \delta^{cd} \del_{\tilde{c}_{\bot}} \del_{\tilde{d}_{\bot}} e^{-2A(p-3)} \nn\\
+ & 2 \delta^{cd} e^{2A(p-3)} \del_{\tilde{c}_{\bot}} e^{-2A(p-3)} \left(  -  e^{2A} \tilde{f}^{a_{||}}{}_{d_{\bot}a_{||}}  + (p-3) \del_{\tilde{d}_{\bot}} e^{2A} \right) \ .\nn
\eea
Generically, $\nabla_a V_b = \del_a V_b - \omega_a{}^c{}_b V_c$ and $\omega_{(a}{}^c{}_{b)}= \delta^{cd} f^e{}_{d(a} \delta_{b)e}$, so we deduce, with $\tilde{f}^{a}{}_{d_{\bot}a}=0$,
\bea
& \tilde{\Delta}_{\bot} e^{-2A(p-3)} = \delta^{cd} \del_{\tilde{c}_{\bot}} \del_{\tilde{d}_{\bot}} e^{-2A(p-3)} + \delta^{cd} \tilde{f}^{a_{||}}{}_{d_{\bot}a_{||}} \del_{\tilde{c}_{\bot}}  e^{-2A(p-3)} \ ,\\
& 4 (\nabla \del \phi)_{6||}= 4 \delta^{AB=a_{||}b_{||}} \nabla_A \del_B \phi =  2 \delta^{cd} e^{2\phi+2A} \tilde{f}^{a_{||}}{}_{d_{\bot}a_{||}} \del_{\tilde{c}_{\bot}} e^{-2\phi} - (p-3) \delta^{cd} e^{2\phi}  \del_{\tilde{c}_{\bot}} e^{-2\phi} \del_{\tilde{c}_{\bot}} e^{2A} \ ,\nn
\eea
where $\tilde{\Delta}_{\bot}$ is the Laplacian on the transverse subspace with smeared metric, i.e.~involving only indices ${}_{\tilde{a}_{\bot}}$. Point \ref{it:pt6} on sources in Section \ref{sec:compactif} requires a compact transverse unwarped subspace without boundaries, implying $\tilde{f}^{a_{\bot}}{}_{d_{\bot}a_{\bot}} = - \tilde{f}^{a_{||}}{}_{d_{\bot}a_{||}} = 0$. Setting this to zero, we eventually obtain
\bea
& \hspace{-0.1in} 2 {\cal R}_{6||} = 2 {\cal R}_{6||}|_{(\del A=0)} + e^{2A(p-2)}\tilde{\Delta}_{\bot} e^{-2A(p-3)} + 2 (p-3) \delta^{cd} e^{2A(p-3)} \del_{\tilde{c}_{\bot}} e^{-2A(p-3)}  \del_{\tilde{d}_{\bot}} e^{2A} \label{Ricciflatpartrace3} \\
& \hspace{-0.1in} \mbox{where}\ \ 2 {\cal R}_{6||}|_{(\del A=0)}  =  2 {\cal R}_{||} + 2 {\cal R}_{||}^{\bot} + \frac{1}{2} \delta^{ch}\delta^{dj}\delta_{ab} f^{a_{||}}{}_{c_{\bot}j_{\bot}} f^{b_{||}}{}_{h_{\bot}d_{\bot}} \ ,\label{R6pardA0}\\
& \hspace{-0.1in} 4 (\nabla \del \phi)_{6||} = - (p-3) \delta^{cd} e^{2\phi}  \del_{\tilde{c}_{\bot}} e^{-2\phi} \del_{\tilde{d}_{\bot}} e^{2A} \ .\label{ndphi6parflat}
\eea
Finally, for a function $f$ such that $\del_{a_{||}} f=0$, one gets using the above
\beq
\Delta_6 f = e^{2A} \tilde{\Delta}_{\bot} f + (p-5) \delta^{ab} \del_{\tilde{a}_{\bot}} e^{2A} \del_{\tilde{b}_{\bot}} f \ .
\eeq
Using this in \eqref{R4} and in \eqref{Deltaphi}, together with \eqref{Ndphi4}, \eqref{Ricciflatpartrace3}, \eqref{ndphi6parflat}, and the dilaton expression \eqref{dilatonwarpap}, we obtain for the quantity \eqref{qtty0ap}
\bea
& 2 {\cal R}_4 + 4 (\nabla\del \phi)_4 - 4 |\del \phi|^2 + 2 \Delta \phi  + 4 (\nabla\del \phi)_{6||} + 2 {\cal R}_{6||} \\
& =  2 e^{-2A} \tilde{{\cal R}}_4 +  2 {\cal R}_{6||}|_{(\del A=0)}  -4 \tilde{\Delta}_{\bot} e^{2A} + 4 e^{-2A} |\widetilde{\d e^{2A}}|^2 \ ,\nn
\eea
with $|\widetilde{\d e^{2A}}|^2 = \delta^{cd} \del_{\tilde{c}_{\bot}} e^{2A} \del_{\tilde{d}_{\bot}} e^{2A}$. It is more convenient to make the quantity $\tilde{\Delta}_{\bot} e^{-4A}$ appear: in our conventions, this quantity is produced by $\d F_k$ in the BI (see e.g.~\cite{Andriot:2015sia} for explicit examples), thus typically generates the $\delta$ functions that localize the $D_p$ and $O_p$ sources, as in $T_{10}$. Then, using $\tilde{\Delta}_{\bot} e^{-4A} = -2 e^{-6A} \tilde{\Delta}_{\bot} e^{2A} + 6 e^{-8A} |\widetilde{\d e^{2A}}|^2 $, we rewrite the above as
\bea
& 2 {\cal R}_4 + 4 (\nabla\del \phi)_4 - 4 |\del \phi|^2 + 2 \Delta \phi  + 4 (\nabla\del \phi)_{6||} + 2 {\cal R}_{6||} \label{usefuleq} \\
& =  2 e^{-2A} \tilde{{\cal R}}_4 +  2 {\cal R}_{6||}|_{(\del A=0)}  +2 e^{6A} \tilde{\Delta}_{\bot} e^{-4A} - 2 e^{10A} |\widetilde{\d e^{-4A}}|^2 \ .\nn
\eea
One can finally make use of \eqref{usefuleq} in \eqref{combine1}, with ${\cal R}_{6||}|_{(\del A=0)}$ given in \eqref{R6pardA0}.

We now detail the following rewriting useful for \eqref{combine1}. One has $e^{2\phi}|F_{k}^{(0)}|^2= e^{10A} |\widetilde{g_s F_{k}^{(0)}}|^2$ thanks to the dilaton \eqref{dilatonwarpap}, and analogously to \eqref{blasquare} with \eqref{coefsmeared},
\bea
|\widetilde{g_s F_{k}^{(0)}}|^2 &= |\varepsilon_p g_s F_{k}^{(0)} - \tilde{*}_{\bot} \d e^{-4A} + \tilde{*}_{\bot} \d e^{-4A} |^{\widetilde{2}} \label{trickap}\\
&= |\varepsilon_p g_s F_{k}^{(0)} - \tilde{*}_{\bot} \d e^{-4A} |^{\widetilde{2}} + |\tilde{*}_{\bot} \d e^{-4A} |^{\widetilde{2}} + 2 \left(\d e^{-4A} \w \left( \varepsilon_p g_s F_{k}^{(0)} - \tilde{*}_{\bot} \d e^{-4A} \right)\right)_{\widetilde{\bot}} \nn \\
&= e^{2\phi -10A}|\varepsilon_p  F_{k}^{(0)} - g_s^{-1} \tilde{*}_{\bot} \d e^{-4A} |^2 + |\widetilde{\d e^{-4A} }|^2 - e^{-12A} \left(\d e^{8A} \w \left( \varepsilon_p g_s F_{k}^{(0)} - \tilde{*}_{\bot} \d e^{-4A} \right)\right)_{\widetilde{\bot}} \nn \\
&= e^{2\phi -10A}| g_s^{-1} \tilde{*}_{\bot} \d e^{-4A} - \varepsilon_p  F_{k}^{(0)} |^2 + |\widetilde{\d e^{-4A} }|^2 + e^{-12A} \left(\d \left(e^{8A} \tilde{*}_{\bot} \d e^{-4A} - e^{8A} \varepsilon_p g_s F_{k}^{(0)} \right)\right)_{\widetilde{\bot}} \nn\\
& - e^{-4A} \left(\d \left( \tilde{*}_{\bot} \d e^{-4A} - \varepsilon_p g_s F_{k}^{(0)}  \right)\right)_{\widetilde{\bot}} \ . \nn
\eea

\end{appendix}


\begin{thebibliography}{100}

\bibitem{Ade:2015tva}
{\bf BICEP2, Planck} Collaboration, P.~Ade {\em et.~al.}, {\it {Joint Analysis
  of BICEP2/Keck Array and Planck Data}},  {\em Phys. Rev. Lett.} {\bf
  114} (2015) 101301 [\href{http://arxiv.org/abs/1502.00612}{{\tt
  arXiv:1502.00612}}].

\bibitem{Ade:2015xua}
{\bf Planck} Collaboration, P.~Ade {\em et al.}, {\it Planck 2015 results. XIII. Cosmological parameters}, {\em Astron. Astrophys.} {\bf 594} (2016) A13 [\href{http://arxiv.org/abs/1502.01589}{{\tt arXiv:1502.01589}}].

\bibitem{Ade:2015ava}
{\bf Planck} Collaboration, P.~Ade {\em et al.}, {\it Planck 2015 results. XVII. Constraints on primordial non-Gaussianity}, {\em Astron. Astrophys.} {\bf 594} (2016) A17 [\href{http://arxiv.org/abs/1502.01592}{{\tt arXiv:1502.01592}}].

\bibitem{Ade:2015lrj}
{\bf Planck} Collaboration, P.~Ade {\em et al.}, {\it Planck 2015 results. XX. Constraints on inflation}, {\em Astron. Astrophys.} {\bf 594} (2016) A20  [\href{http://arxiv.org/abs/1502.02114}{{\tt arXiv:1502.02114}}].

\bibitem{Chen:2016vvw}
X.~Chen, C.~Dvorkin, Z.~Huang, M.~H.~Namjoo and L.~Verde, {\it The Future of Primordial Features with Large-Scale Structure Surveys}, {\em JCAP} {\bf 11} (2016) 014 [\href{http://arxiv.org/abs/1605.09365}{{\tt arXiv:1605.09365}}].

\bibitem{Parameswaran:2016fqr}
S.~L.~Parameswaran and I.~Zavala, {\it Prospects for Primordial Gravitational Waves in String Inflation}, {\em Int. J. Mod. Phys.} {\bf D 25} (2016) 12 1644011 [\href{http://arxiv.org/abs/1606.02537}{{\tt arXiv:1606.02537}}].

\bibitem{Conlon:2006tq}
J.~P.~Conlon, {\it The QCD axion and moduli stabilisation}, {\em JHEP} {\bf 05} (2006) 078 [\href{http://arxiv.org/abs/hep-th/0602233}{{\tt hep-th/0602233}}].

\bibitem{Andriot:2015aza}
D.~Andriot, {\it A no-go theorem for monodromy inflation}, {\em JCAP} {\bf 03} (2016) 025 [\href{http://arxiv.org/abs/1510.02005}{{\tt arXiv:1510.02005}}].

\bibitem{Kachru:2003aw}
S.~Kachru, R.~Kallosh, A.~D.~Linde and S.~P.~Trivedi, {\it De Sitter vacua in string theory}, {\em Phys. Rev.} {\bf D 68} (2003) 046005 [\href{http://arxiv.org/abs/hep-th/0301240}{{\tt hep-th/0301240}}].

\bibitem{Bena:2009xk}
I.~Bena, M.~Gra\~na and N.~Halmagyi, {\it On the Existence of Meta-stable Vacua in Klebanov-Strassler}, {\em JHEP} {\bf 09} (2010) 087 [\href{https://arxiv.org/abs/0912.3519}{{\tt arXiv:0912.3519}}].

\bibitem{Bena:2016fqp}
I.~Bena, J.~Bl{\aa}b{\"a}ck and D.~Turton, {\it Loop corrections to the antibrane potential}, {\em JHEP} {\bf 07} (2016) 132 [\href{https://arxiv.org/abs/1602.05959}{{\tt arXiv:1602.05959}}].

\bibitem{deAlwis:2016cty}
S.~P.~de Alwis, {\it Constraints on Dbar Uplifts}, {\em JHEP} {\bf 11} (2016) 045 [\href{http://arxiv.org/abs/1605.06456}{{\tt arXiv:1605.06456}}].

\bibitem{CaboBizet:2016qsa}
N.~Cabo Bizet and S.~Hirano, {\it Revisiting constraints on uplifts to de Sitter vacua}, [\href{http://arxiv.org/abs/1607.01139}{{\tt arXiv:1607.01139}}].

\bibitem{Green:2011cn}
S.~R.~Green, E.~J.~Martinec, C.~Quigley and S.~Sethi, {\it Constraints on String Cosmology}, {\em Class. Quant. Grav.} {\bf 29} (2012) 075006 [\href{http://arxiv.org/abs/1110.0545}{{\tt arXiv:1110.0545}}].

\bibitem{Gautason:2012tb}
F.~F.~Gautason, D.~Junghans and M.~Zagermann, {\it On  Cosmological Constants from alpha'-Corrections}, {\em JHEP} {\bf 06} (2012) 029 [\href{http://arxiv.org/abs/1204.0807}{{\tt arXiv:1204.0807}}].

\bibitem{Kutasov:2015eba}
D.~Kutasov, T.~Maxfield, I.~Melnikov and S.~Sethi, {\it Constraining de Sitter Space in String Theory}, {\em Phys. Rev. Lett.} {\bf 115} (2015) 071305 [\href{http://arxiv.org/abs/1504.00056}{{\tt arXiv:1504.00056}}].

\bibitem{Quigley:2015jia}
C.~Quigley, {\it Gaugino Condensation and the Cosmological Constant}, {\em JHEP} {\bf 06} (2015) 104 [\href{http://arxiv.org/abs/1504.00652}{{\tt arXiv:1504.00652}}].

\bibitem{Gibbons:1984kp}
G.~W.~Gibbons, {\it Aspects of supergravity theories}, {\em XV GIFT Seminar on Supersymmetry and Supergravity} (1984).

\bibitem{deWit:1986xg}
B.~de Wit, D.~Smit and N.~Hari Dass, {\it Residual Supersymmetry of Compactified D=10 Supergravity}, {\em Nucl. Phys.} {\bf B 283} (1987) 165.

\bibitem{Maldacena:2000mw}
J.~M.~Maldacena and C.~Nu\~nez, {\it Supergravity description of field theories on curved manifolds and a no go theorem}, {\em Int. J. Mod. Phys.} {\bf A 16} (2001) 822 [\href{http://arxiv.org/abs/hep-th/0007018}{{\tt hep-th/0007018}}].

\bibitem{Townsend:2003qv}
P.~K.~Townsend, {\it Cosmic acceleration and M theory}, [\href{http://arxiv.org/abs/hep-th/0308149}{{\tt hep-th/0308149}}].

\bibitem{Hertzberg:2007wc}
M.~P.~Hertzberg, S.~Kachru, W.~Taylor and M.~Tegmark, {\it Inflationary Constraints on Type IIA String Theory}, {\em JHEP} {\bf 12} (2007) 095 [\href{http://arxiv.org/abs/0711.2512}{{\tt arXiv:0711.2512}}].

\bibitem{Haque:2008jz}
S.~S.~Haque, G.~Shiu, B.~Underwood and T.~Van Riet, {\it Minimal simple de Sitter solutions}, {\em Phys. Rev.} {\bf D 79} (2009) 086005 [\href{http://arxiv.org/abs/0810.5328}{{\tt arXiv:0810.5328}}].

\bibitem{Caviezel:2008tf}
C.~Caviezel, P.~Koerber, S.~Kors, D.~L\"ust, T.~Wrase and M.~Zagermann, {\it On the Cosmology of Type IIA Compactifications on SU(3)-structure Manifolds}, {\em JHEP} {\bf 04} (2009) 010 [\href{http://arxiv.org/abs/0812.3551}{{\tt arXiv:0812.3551}}].

\bibitem{Flauger:2008ad}
R.~Flauger, S.~Paban, D.~Robbins and T.~Wrase, {\it Searching for slow-roll moduli inflation in massive type IIA supergravity with metric fluxes}, {\em Phys. Rev.} {\bf D 79} (2009) 086011 [\href{http://arxiv.org/abs/0812.3886}{{\tt arXiv:0812.3886}}].

\bibitem{Danielsson:2009ff}
U.~H.~Danielsson, S.~S.~Haque, G.~Shiu and T.~Van Riet, {\it Towards Classical de Sitter Solutions in String Theory}, {\em JHEP} {\bf 09} (2009) 114 [\href{http://arxiv.org/abs/0907.2041}{{\tt arXiv:0907.2041}}].

\bibitem{deCarlos:2009fq}
B.~de Carlos, A.~Guarino and J.~M.~Moreno, {\it Flux moduli stabilisation, Supergravity algebras and no-go theorems}, {\em JHEP} {\bf 01} (2010) 012 [\href{http://arxiv.org/abs/0907.5580}{{\tt arXiv:0907.5580}}].

\bibitem{Caviezel:2009tu}
C.~Caviezel, T.~Wrase and M.~Zagermann, {\it Moduli Stabilization and Cosmology of Type IIB on SU(2)-Structure Orientifolds}, {\em JHEP} {\bf 04} (2010) 011 [\href{http://arxiv.org/abs/0912.3287}{{\tt arXiv:0912.3287}}].

\bibitem{Wrase:2010ew}
T.~Wrase and M.~Zagermann, {\it On Classical de Sitter Vacua in String Theory}, {\em Fortsch. Phys.} {\bf 58} (2010) 906 [\href{http://arxiv.org/abs/1003.0029}{{\tt arXiv:1003.0029}}].

\bibitem{Burgess:2011rv}
C.~P.~Burgess, A.~Maharana, L.~van Nierop, A.~A.~Nizami and F.~Quevedo, {\it On Brane Back-Reaction and de Sitter Solutions in Higher-Dimensional Supergravity}, {\em JHEP} {\bf 04} (2012) 018 [\href{http://arxiv.org/abs/1109.0532}{{\tt arXiv:1109.0532}}].

\bibitem{VanRiet:2011yc}
T.~Van Riet, {\it On classical de Sitter solutions in higher dimensions}, {\em Class. Quant. Grav.} {\bf 29} (2012) 055001 [\href{http://arxiv.org/abs/1111.3154}{{\tt arXiv:1111.3154}}].

\bibitem{Gautason:2013zw}
F.~F.~Gautason, D.~Junghans and M.~Zagermann, {\it Cosmological Constant, Near Brane Behavior and Singularities}, {\em JHEP} {\bf 09} (2013) 123 [\href{http://arxiv.org/abs/1301.5647}{{\tt arXiv:1301.5647}}].

\bibitem{Kallosh:2014oja}
R.~Kallosh, A.~Linde, B.~Vercnocke and T.~Wrase, {\it Analytic Classes of Metastable de Sitter Vacua}, {\em JHEP} {\bf 10} (2014) 011 [\href{http://arxiv.org/abs/1406.4866}{{\tt arXiv:1406.4866}}].

\bibitem{Silverstein:2007ac}
E.~Silverstein, {\it Simple de Sitter Solutions}, {\em Phys. Rev.} {\bf D 77} (2008) 106006 [\href{http://arxiv.org/abs/0712.1196}{{\tt arXiv:0712.1196}}].

\bibitem{Blaback:2010sj}
J.~Bl{\aa}b\"ack, U.~H.~Danielsson, D.~Junghans, T.~Van Riet, T.~Wrase and M.~Zagermann, {\it Smeared versus localised sources in flux compactifications}, {\em JHEP} {\bf 12} (2010) 043 [\href{http://arxiv.org/abs/1009.1877}{{\tt arXiv:1009.1877}}].

\bibitem{Junghans:2013xza}
D.~Junghans, {\it Backreaction of Localised Sources in String Compactifications}, [\href{http://arxiv.org/abs/1309.5990}{{\tt arXiv:1309.5990}}].

\bibitem{Danielsson:2010bc}
U.~H.~Danielsson, P.~Koerber and T.~Van Riet, {\it Universal de Sitter solutions at tree-level}, {\em JHEP} {\bf 05} (2010) 090 [\href{http://arxiv.org/abs/1003.3590}{{\tt arXiv:1003.3590}}].

\bibitem{Danielsson:2011au}
U.~H.~Danielsson, S.~S.~Haque, P.~Koerber, G.~Shiu, T.~Van Riet and T.~Wrase, {\it De Sitter hunting in a classical landscape}, {\em Fortsch. Phys.} {\bf 59} (2011) 897 [\href{http://arxiv.org/abs/1103.4858}{{\tt arXiv:1103.4858}}].

\bibitem{Covi:2008ea}
L.~Covi, M.~Gomez-Reino, C.~Gross, J.~Louis, G.~A.~Palma and C.~A.~Scrucca, {\it de Sitter vacua in no-scale supergravities and Calabi-Yau string models}, {\em JHEP} {\bf 06} (2008) 057 [\href{http://arxiv.org/abs/0804.1073}{{\tt arXiv:0804.1073}}].

\bibitem{Shiu:2011zt}
G.~Shiu and Y.~Sumitomo, {\it Stability Constraints on Classical de Sitter Vacua}, {\em JHEP} {\bf 09} (2011) 052 [\href{http://arxiv.org/abs/1107.2925}{{\tt arXiv:1107.2925}}].

\bibitem{Danielsson:2012et}
U.~H.~Danielsson, G.~Shiu, T.~Van Riet and T.~Wrase, {\it A note on obstinate tachyons in classical dS solutions}, {\em JHEP} {\bf 03} (2013) 138 [\href{http://arxiv.org/abs/1212.5178}{{\tt arXiv:1212.5178}}].

\bibitem{Junghans:2016uvg}
D.~Junghans, {\it Tachyons in Classical de Sitter Vacua}, {\em JHEP} {\bf 06} (2016) 132, [\href{http://arxiv.org/abs/1603.08939}{{\tt arXiv:1603.08939}}].

\bibitem{Giddings:2001yu}
S.~B.~Giddings, S.~Kachru and J.~Polchinski, {\it Hierarchies from Fluxes in String Compactifications}, {\em Phys. Rev.} \textbf{D 66} (2002) 106006 [\href{http://arxiv.org/abs/hep-th/0105097}{{\tt hep-th/0105097}}].

\bibitem{Andriot:2016ufg}
D.~Andriot, J.~Bl{\aa}b{\"a}ck and T.~Van Riet, {\it Minkowski flux vacua of type II supergravities}, {\em Phys. Rev. Lett.} {\bf 118} (2017) 1 011603 [\href{http://arxiv.org/abs/1609.00729}{{\tt arXiv:1609.00729}}].

\bibitem{Andriot:2010ju}
D.~Andriot, E.~Goi, R.~Minasian and M.~Petrini, {\it {Supersymmetry breaking branes on solvmanifolds and de Sitter vacua in string theory}},  {\em JHEP} {\bf 05} (2011) 028 [\href{http://arxiv.org/abs/1003.3774}{{\tt arXiv:1003.3774}}].

\bibitem{Andriot:2015sia}
D.~Andriot, {\it New supersymmetric vacua on solvmanifolds}, {\em JHEP} {\bf 02} (2016) 112 [\href{http://arxiv.org/abs/1507.00014}{{\tt arXiv:1507.00014}}].

\bibitem{Martucci:2005ht}
L.~Martucci and P.~Smyth, {\it Supersymmetric $D$-branes and calibrations on general ${\cal N}=1$ backgrounds}, {\em JHEP} \textbf{11} (2005) 048 [\href{http://arxiv.org/abs/hep-th/0507099}{{\tt hep-th/0507099}}].

\bibitem{Villadoro:2007tb}
G.~Villadoro and F.~Zwirner, {\it On general flux backgrounds with localized sources}, {\em JHEP} {\bf 11} (2007) 082 [\href{http://arxiv.org/abs/0710.2551}{{\tt arXiv:0710.2551}}].

\bibitem{Shelton:2005cf}
J.~Shelton, W.~Taylor and B.~Wecht, {\it Nongeometric flux compactifications}, {\em JHEP} {\bf 10} (2005) 085 [\href{http://www.arXiv.org/abs/hep-th/0508133}{{\tt hep-th/0508133}}].

\bibitem{Silverstein:2008sg}
E.~Silverstein and A.~Westphal, {\it Monodromy in the CMB: gravity waves and string inflation}, {\em Phys. Rev.} {\bf D 78} (2008) 106003 [\href{http://arxiv.org/abs/0803.3085}{{\tt arXiv:0803.3085}}].

\bibitem{Koerber:2007jb}
P.~Koerber and L.~Martucci, {\it D-branes on AdS flux compactifications}, {\em JHEP} {\bf 01} (2008) 047 [\href{http://arxiv.org/abs/0710.5530}{{\tt arXiv:0710.5530}}].

\bibitem{Berasaluce-Gonzalez:2016kqb}
M.~Berasaluce-Gonz\'alez, G.~Honecker and A.~Seifert, {\it Towards Geometric D6-Brane Model Building on non-Factorisable Toroidal Z4-Orbifolds}, {\em JHEP} {\bf 08} (2016) 062 [\href{http://arxiv.org/abs/1606.04926}{{\tt arXiv:1606.04926}}].





\bibitem{Andriot:2011iw}
D.~Andriot, {\it Heterotic string from a higher dimensional perspective}, {\em Nucl. Phys.} {\bf B 855} (2012) 222 [\href{http://arxiv.org/abs/1102.1434}{{\tt arXiv:1102.1434}}].

\bibitem{Bergshoeff:2001pv}
E.~Bergshoeff, R.~Kallosh, T.~Ortin, D.~Roest and A.~Van Proeyen, \textit{New formulations of D = 10 supersymmetry and D8 - O8 domain walls}, {\em Class. Quant. Grav.} \textbf{18} (2001) 3359 [\href{http://arxiv.org/abs/hep-th/0103233}{{\tt hep-th/0103233}}].

\bibitem{Koerber:2007hd}
P.~Koerber and D.~Tsimpis, {\it Supersymmetric sources, integrability and generalized-structure compactifications}, {\em JHEP} \textbf{08} (2007) 082 [\href{http://arxiv.org/abs/0706.1244}{{\tt arXiv:0706.1244}}].

\bibitem{Grana:2006kf}
M.~Gra\~na, R.~Minasian, M.~Petrini and A.~Tomasiello, {\it {A Scan for new N=1 vacua on twisted tori}},  {\em JHEP} {\bf 05} (2007) 031 [\href{http://arxiv.org/abs/hep-th/0609124}{{\tt hep-th/0609124}}].

\bibitem{Bergshoeff:1996tu}
E.~Bergshoeff and P.~K.~Townsend, {\it Super D-branes}, {\em Nucl. Phys.} {\bf B 490} (1997) 145 [\href{http://arxiv.org/abs/hep-th/9611173}{{\tt hep-th/9611173}}].

\bibitem{Lust:2008zd}
D.~L\"ust, F.~Marchesano, L.~Martucci and D.~Tsimpis, {\it Generalized non-supersymmetric flux vacua}, {\em JHEP} {\bf 11} (2008) 021 [\href{http://arxiv.org/abs/0807.4540}{{\tt arXiv:0807.4540}}].

\bibitem{Koerber:2010bx}
P.~Koerber, {\it Lectures on Generalized Complex Geometry for Physicists}, {\em Fortsch. Phys.} {\bf 59} (2011) 169 [\href{http://arxiv.org/abs/1006.1536}{{\tt arXiv:1006.1536}}].

\bibitem{Koerber:2005qi}
P.~Koerber, {\it Stable D-branes, calibrations and generalized Calabi-Yau geometry}, {\em JHEP} \textbf{08} (2005) 099 [\href{http://arxiv.org/abs/hep-th/0506154}{{\tt hep-th/0506154}}].

\bibitem{Koerber:2006hh}
P.~Koerber and L.~Martucci, \textit{Deformations of calibrated D-branes in flux generalized complex manifolds}, {\em JHEP} \textbf{12} (2006) 062 [\href{http://arxiv.org/abs/hep-th/0610044}{{\tt hep-th/0610044}}].

\bibitem{Witt}
F.~Gmeiner and F.~Witt, {\it Calibrated cycles and T-duality}, {\em Commun. Math. Phys.} {\bf 283} (2008) 543 [\href{http://arxiv.org/abs/math/0605710}{{\tt math/0605710}}].

\bibitem{Martucci:2011dn}
L.~Martucci, {\it Electrified branes}, {\em JHEP} {\bf 02} (2012) 097 [\href{http://arxiv.org/abs/1110.0627}{{\tt arXiv:1110.0627}}].

\bibitem{Andriot:2013xca}
D.~Andriot and A.~Betz, {\it $\beta$-supergravity: a ten-dimensional theory with non-geometric fluxes, and its geometric framework}, {\em JHEP} \textbf{12} (2013) 083 [\href{http://arxiv.org/abs/1306.4381}{{\tt arXiv:1306.4381}}].

\bibitem{Andriot:2014uda}
D.~Andriot and A.~Betz, {\it NS-branes, source corrected Bianchi identities, and more on backgrounds with non-geometric fluxes},  {\em JHEP} {\bf 07} (2014) 059 [\href{http://arxiv.org/abs/1402.5972}{{\tt arXiv:1402.5972}}].



\end{thebibliography}
\end{document}